\documentclass[twocolumn,letterpaper,amsmath,amssymb]{revtex4}
\usepackage{graphicx}
\usepackage{dcolumn}
\usepackage{bm}
\usepackage{color}

\newcommand{\bftheta}{ \mbox{\boldmath{$\theta$}}}

\newcommand{\beq}[1]{  \begin{equation} \label{#1} }
\newcommand{\beqa}[1]{\begin{eqnarray} \label{#1} }

\newcommand{\eeq}{\end{equation}}
\newcommand{\eeqa}{\end{eqnarray}}

\begin{document}

\newcommand{\rf}[1]{(\ref{#1})}

\newcommand{\bfomega}{ \mbox{\boldmath{$\omega$}}}

\title{
Dynamic effective mass of granular media and\\ the attenuation of
structure-borne sound}

\author{John Valenza$^1$, Chaur-Jian Hsu$^1$, Rohit Ingale$^{1,2}$, Nicolas Gland$^2$, Hern\'an A. Makse$^2$,
 and David Linton Johnson$^1$}

\affiliation{
$^1$ Schlumberger-Doll Research, One Hampshire Street, Cambridge, MA 02139\\
$^2$ Levich Institute and Physics Department,
City College of New York, New York, NY 10031.
}
\date{\today}

\begin{abstract}

We report a theoretical and experimental investigation into the
fundamental physics of why loose granular media are effective
deadeners of structure-borne sound.  Here we demonstrate that a
measurement of the effective mass, $\tilde{M}(\omega)$, of the
granular medium is a sensitive and direct way to answer the
question: What is the specific mechanism whereby acoustic energy is
transformed into heat?  Specifically, we apply this understanding to
the case of the flexural resonances of a rectangular bar with a
grain-filled cavity within it.  The pore space in the granular
medium is air of varying humidity.  The dominant features of
$\tilde{M}(\omega)$ are a sharp resonance and a broad background,
which we analyze within the context of simple models.  We find that:
a) On a fundamental level, dampening of acoustic modes is dominated
by adsorbed films of water at grain-grain contacts, not by global
viscous dampening or by attenuation within the grains. b)  These
systems may be understood, qualitatively, in terms of a
height-dependent and diameter-dependent effective sound speed [$\sim
100-300 \;(m\cdot s^{-1})$] and an effective viscosity [$\sim 5\times 10^4$
Poise]. c) There is an acoustic Janssen effect in the sense that, at
any frequency, and depending on the method of sample preparation,
approximately one-half of the effective mass is borne by the side
walls of the cavity and one-half by the bottom. d) There is a
monotonically increasing effect of humidity on the dampening of the
fundamental resonance within the granular medium which translates to
a non-monotonic, but predictable, variation of dampening within the
grain-loaded bar.
 \linebreak
\linebreak{PACS Numbers: 45.70.-n, 46.40.-f, 81.05.Rm}
\end{abstract}

\maketitle


\section{Introduction}
Loose grains, made of a variety of different materials, dampen
structure-borne acoustic signals very efficiently when they
partially fill cavities within the structure itself \cite{Kuhl,
cremer, Bourinet1}. For this reason, there is a practical motivation
to develop an effective method to optimize the dampening of unwanted
structure-borne acoustic signals.  The fundamental origins of the
dissipation mechanisms in granular materials are still unknown,
however, making it difficult to optimize the effect. Partly, this is
because, until now, it has not been possible to study the relevant
properties of the granular medium directly, independent of the
structure whose acoustic properties are being modified by the
granular medium.  In this article, we will answer the questions:
Where does the acoustic energy go when it is attenuated by granular
media?  What is the specific microscopic mechanism?

Here we pursue the concept of the effective mass,
$\tilde{M}(\omega)$, of a loose granular aggregate contained within
a rigid cavity \cite{Kurtze, Bourinet2}; a preliminary report on
some of our results has been published previously \cite{Hsu}.
$\tilde{M}(\omega)$ is determined by simultaneously measuring the
force, $F(\omega)$, on the cavity and its acceleration, $a(\omega)$,
when the cavity undergoes oscillation at a frequency $\omega$.  This
measurement allows us to focus very directly on those properties of
the granular medium which affect the propagation and attenuation of
structure-borne sound.  Specifically, if an identically filled
cavity is located within an acoustically resonant structure, we
demonstrate how to predict the changes in the acoustic properties of
the structure, such as sound attenuation or resonance-frequency
shift, based on a knowledge of $\tilde{M}(\omega)$.  This fact
allows us to focus our efforts on understanding the relevant
properties of the granular medium directly, rather than having to
infer its properties via its effects on the host structure.  In this
regard we demonstrate the dominant effect of humidity, both on
$\tilde{M}(\omega)$ and on the acoustic resonance of a steel bar
having a grain-filled cavity.

There have been several previous investigations into the origins of
particle damping.  Cremer and Heckl \cite{cremer} concluded that
damping is especially high when the (vertical) thickness of a
granular layer is equal to an odd  multiple of a quarter wavelength
in the granular medium.  Sun {\it et al} \cite{Sun} have treated the
acoustic effect of the granular medium as if it was due to radiative
damping; they computed the acoustic loss due to radiation by
assuming the granular medium is a low-velocity fluid.  Bourinet and
Le Hou\`{e}dec \cite{Bourinet1} and Varanasi, {\it et al.}
\cite{Varanasi_a,Varanasi_b} have each considered the acoustic
propagation characteristics of long hollow tubes, partially filled
with granular material.  Each approximated the medium as a low
velocity, high attenuation fluid and each achieved quite reasonable
agreement between their computed and their measured values. (The two
theories are similar but differ in their details.) We do not dispute
these aforementioned results which are, in fact, quite reasonable.
We point out, though, that the relevant granular medium parameters
(the sound speed, the loss factor) are generally set by requiring a
match to the observed acoustic characteristics of the grain-loaded
structure. (An exception is Ref.\cite{Varanasi_b}.) It is this last
feature that we obviate in the present article:  For the kinds of
structures we consider here, acoustic loss is determined by the
imaginary component of the effective mass, $M_2(\omega)$, evaluated
at the propagation frequency  (or resonant frequency, as the case
may be). Moreover, we establish that the properties of granular
media are very much dependent on the filling level in the cavity and
we demonstrate that the side walls of the cavity hold up some of the
dynamic load.  Thus the granular material cannot be idealized, for
acoustic purposes, as a fluid, and certainly not a homogeneous one.

Intuitively one might expect that the effect of the granular loading
would be to lower the resonance frequency of the structure holding
the grains.  While this often happens we show situations in which
the real part of $\tilde{M}(\omega)$ is negative in the frequency
range of interest, leading to an {\it increase} in the resonant
frequency of the structure containing the grains.  This behavior has
been observed before by Kang {\it et al} \cite{Kang} who monitored
the resonances in a clamped plate as more and more grains were
loaded on top of it. Initially, as grains are added, the resonance
frequency drops; it reaches a minimum, then increases, eventually
often exceeding the original (unloaded) resonance frequency of the
plate. This behavior has a simple understanding in terms of a
resonance within the grains, whereby the real part of
$\tilde{M}(\omega)$ can take on negative values.  (See Section
\ref{sec:granmdehum}, below.)

Generally speaking, we find that $\tilde{M}(\omega)$ exhibits a
sharp resonance, which, as one part of our analysis, we interpret in
terms of an effective sound speed, albeit one which is dependent
upon the depth of filling of the cavity, and a broad tail that
decreases roughly as $\omega^{-1/2}$, which we interpret in terms of
an effective viscosity.  These general features have been observed
previously, by others \cite{Kurtze,Bourinet2} using similar
measurements.  It is to be emphasized that our experiments are all
done in the regime of linear (small amplitude) acoustics; this
viscosity is not relevant to a granular flow, e.g. in a pipe or in a
loading hopper.  Each of our two interpretations is based on toy
models, which assume that the entire effective mass is borne by the
bottom of the cavity or by the walls, respectively. A concrete
example of the former behavior is provided by the effective mass of
simple liquids; we demonstrate how our technique enables us to
measure the density and the sound speed of four different liquids.

We have developed molecular dynamic simulations to analyze the
expected behavior of $\tilde{M}(\omega)$ under the assumption that
the contacts are described by dampened Hertz-Mindlin theory, with or
without possible global dampening due to the viscosity of the air.
We have found that there is an acoustic Janssen effect in the sense
that, at any frequency, approximately one-half of the effective mass
rests on the bottom of the cavity and one-half on the side walls.
Thus, these toy models have only qualitative validity.
Notwithstanding, it is reasonable to use them to extract approximate
values for the sound speed and the viscosity of our granular
ensembles. Finally, our simulations as well as our experiments on
the effects of humidity on $\tilde{M}(\omega)$ indicate that the
dominant microscopic mechanism for dampening is at the grain-grain
contact level and is not due to global viscous dampening.

We show that this dampening is much larger for high humidity systems
than for low ones leading us to conclude that the mechanism is
related to the viscosity of the adsorbed water in the region of the
contacts.  Such a conclusion accords with the finding in ``room-dry"
sedimentary rocks that acoustic attenuation is caused by
stress-induced diffusion of adsorbed layers of volatile molecules
\cite{Tittmann} and, more recently, by direct measurements of
acoustic attenuation in dry and weakly wet granular media
\cite{Brunet}.

The organization of the paper is as follows: We describe our
experimental technique in Section \ref{sec:expproc}.  There are two
types of measurements here.  In addition to measuring
$\tilde{M}(\omega)$ we also measure the resonant frequencies and
dampening rates of flexural normal modes in a rectangular bar having
a grain-filled cavity in it.   We show how the measured function
$\tilde{M}(\omega)$ can be used to compute the effect of the
granular medium on the resonant frequency and dampening
characteristics of a structure in Section \ref{sec:perth}.  This is
illustrated schematically in Figure \rf{cupbar_scheme}.  When the
cavity in the bar is filled with grains its own resonance frequency
is changed from $f_0$ in the unloaded state to $f_R$, which is
complex-valued, reflecting the attenuation in the problem.
Additionally, the effective mass, $\tilde{M}(\omega)$, generally has
its own (complex-valued) resonance frequencies, $f_g$.  It may
happen that one or more of these resonances is manifest as
subsidiary resonances within the grain-loaded bar system, although
the complex-valued resonance frequency is changed by virtue of the
bar's compliance.

\begin{figure}
\hbox{ \resizebox{6.0cm}{!} {
\includegraphics{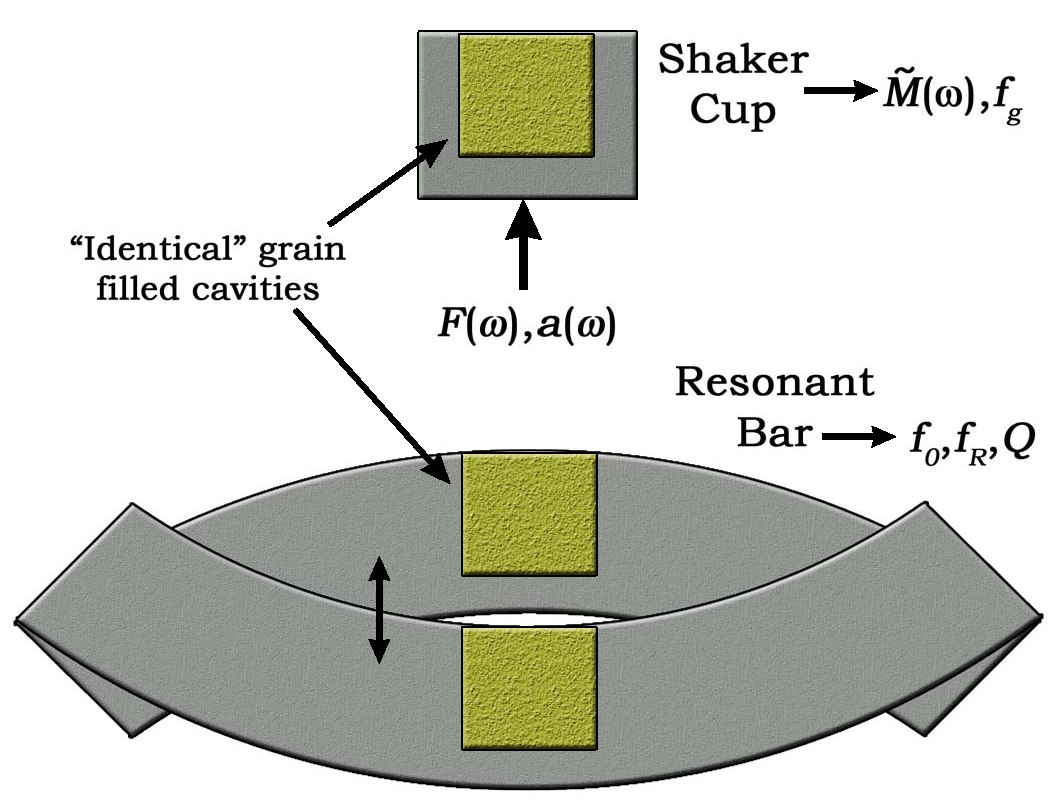}} }
\caption {(Color online) Schematic of measurements:  The effective
mass, $\tilde{M}(\omega)$, is determined by simultaneously measuring
the acceleration, $a(\omega)$, and force, $F(\omega)$, on an
oscillating cup filled with grains.  Using the measured
$\tilde{M}(\omega)$ we show how to predict the acoustic properties
of a resonance structure having the same grain-filled cavity.  Here,
$f_g$ represents a resonance frequency within the granular medium,
$f_0$ represents the resonance frequency of the unloaded structure,
and $f_R$ represents the resonance frequency of the loaded
structure.  In general, these resonance frequencies are
complex-valued.}\label{cupbar_scheme}
\end{figure}

Next, in Section \ref{sec:genprop} we discuss some rather general
properties of the dynamic effective mass, considered as a causal
response function.  We analyze a model in which we treat the
granular medium as a collection of rigid objects interacting via
contact forces between them.  We also analyze simple, continuum
mechanical, models.  Section \ref{sec:macprop} is devoted to an
analysis of our data in the context of these continuum models which
are useful for understanding features such as the main resonance,
$f_g$, and the high frequency tail. Here, we analyze data on
cavities filled with simple liquids, as well as data on our granular
media. In order to get a sense of whether contact dampening or
global dampening is the dominant mechanism in granular media we have
performed a series of numerical simulations, which we report in
Section \ref{sec:simulations}.  In Section \ref{sec:timo} we
investigate the effect of differing humidities both on
$\tilde{M}(\omega)$ and on the flex bar resonances.  By means of
these data it becomes clear that the dampening is local, due to the
viscous adsorbed films of water in the contact region.  We summarize
our conclusions in Section \ref{sec:conclusions}.

\section{Experimental Procedure}\label{sec:expproc}
\subsection{Effective Mass of Granular Media}
A cylindrical cavity (of diameter 2.54 cm and height 3.07 cm)
excavated in a rigid Al cup is filled with tungsten particles. Each
of these particles consists of four or five equal-axis particles, of
nominal size 100 $\mu$m, fused together. (See Figure
\ref{W_granules_photo}, below.)  The individual grains are far from
being identical; using a micro-balance we measured the individual
masses of 19 of these grains from which we deduce the average mass
of an individual grain to be $m_g = 44.2 \pm 18.0\:\mu$g.  The large
density of tungsten maximizes the effects we are studying. The cup
is subjected to a vertical sinusoidal vibration at angular frequency
$\omega$; the resulting acceleration is measured with either one or
two accelerometers attached to the cup, on the underside to one side
of center (see below), and the force is measured with a force gauge
mounted between the shaker and the cup. Taking into account the mass
of the empty cup, $M_c$, we have \beq{def} \tilde{M}(\omega) + M_c =
\frac{F(\omega)}{a(\omega)},
\end{equation}
where the effective mass of the granular medium, $\tilde{M}(\omega)
= M_1(\omega) \:+\: iM_2(\omega)$, is complex-valued, reflecting the
partially in-phase, partially out-of-phase motion of the individual
grains, relative to the cup motion.  As we shall see, $M_1$ may be
conceptualized as an inertial effect, in the conventional sense,
except that it depends upon an interplay between the masses of the
individual grains and the stiffnesses of the grain-grain contacts;
$M_2$ describes the attenuative aspects of the medium.

\begin{figure}
\hbox{ \resizebox{8.5cm}{!} {
\includegraphics{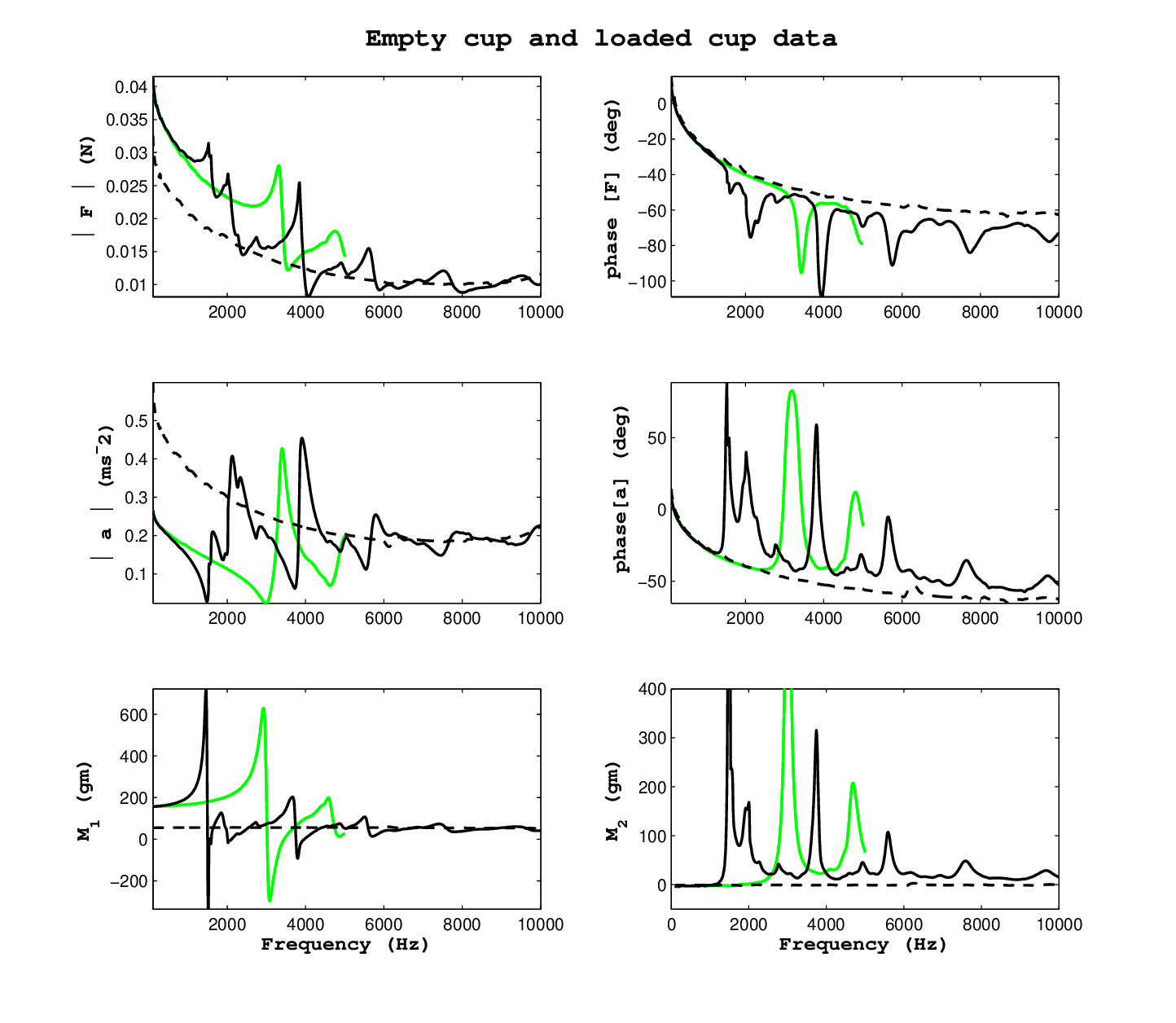}} }
\caption {(Color online) Measurement results using one specific
shaker cup over a wide frequency range. The top row represents the
measured force, the middle row the acceleration, and the bottom row
the effective mass. The left column represents the magnitude and the
right column the phase relative to the driving voltage on the
shaker, except that the bottom row shows the real, $M_1$, and the
imaginary, $M_2$, parts of $\tilde{M}(\omega)$. The dashed curves
represent the empty cup.  The black curves are for a cup that was
filled by simply tapping it whereas the green represents the
mechanical compaction protocol, as described in the text.   The
measured effective mass of the empty cup, shown as a dashed curve,
is essentially a frequency independent constant, which constant is
subtracted from the total effective mass to yield that of the
granular medium.}\label{F_a_m}
\end{figure}

Of course, there is no such thing as a perfectly rigid cup. We have
used a variety of different cups of slightly different geometries,
with the intention that the apparent effect mass of the empty cup
should be approximately frequency-independent. In Figure \ref{F_a_m}
we show the results for one such cup plotted over a fairly wide
frequency range.  We show both the empty cup data (dashed lines) and
the filled cup data (solid lines).  Depending upon the way in which
the cup is filled with grains, either by slightly tapping on the cup
(black curve) or by mechanically compacting the grains with press
and plunger (green curve) we get quite different results, as
discussed below.  We note that the effective mass of the empty cup
is essentially a frequency-independent constant, as expected.  This
result is a consequence of our use of two accelerometers on either
side of the bottom of the cup, and taking the average.  In either
accelerometer there is a visible resonance structure around 6 (kHz),
which seems to be a consequence of a ``wobble" motion; the cup does
not, literally, oscillate along a vertical axis.  By averaging the
two accelerometer signals, this wobble motion is effectively
canceled. In this way we have improved on the technique reported in
Ref. \cite{Hsu}.

We note from Figure \ref{F_a_m} that in the low frequency limit,
$\tilde M(\omega)$ tends to the static mass of the grains.   Also,
there is a relatively sharp resonance peak whose position depends
strongly on the manner in which the grains were prepared;  there are
also subsidiary resonances at higher frequencies. Although the data
in Figure \ref{F_a_m} are for the tungsten grains, we have shown
previously that these general features of $\tilde{M}(\omega)$ are
present in other granular media, such as spherical glass beads or
spherical lead beads  \cite{Hsu}. In this article we focus on the
tungsten granules, simply because the effect is maximal for such
dense particles.

\subsection{Resonant Bar}\label{sec:resbar}

We consider the resonant frequencies of a rectangular bar of
stainless steel whose dimensions are L X W X H = 20.32 cm X 3.81 cm
X 3.18 cm.  In the center of the top surface a cavity is excavated
having virtually identical dimensions as that in the shaker cup
(Figure \ref{cupbar_scheme}).  This cavity is filled with tungsten
granules in the same manner and with the same mass as that held in
the shaker.

We monitor the flexural modes of the system:  The bar is suspended
by wire supports attached at the approximate locations of the
displacement nodes of the fundamental flex mode of the bar [Table
3.2 of reference \cite{Kinsler}].  The purpose here is to minimize
any additional dampening in the experiment due to radiation of
energy into the bar supports.  On the top and on the bottom of the
bar, near each of the ends, we epoxy piezoceramic disks of 6.35 mm
diameter and 3.18 mm thickness.  The two top disks are driven in
phase with each other while the bottom two are driven out of phase
with those on top.  In this configuration the four disks act as a
bending moment on the bar, thus inducing the desired flexural
motion.  We mount an accelerometer on the bottom directly under the
center of the cavity.  The bending moment is driven at a constant
voltage as the frequency is swept in increments which are finely
spaced when the Q is large (as for an empty cavity) and more
coarsely spaced when the Q is low, as when the cavity is grain
loaded.  The output of the accelerometer, $a(\omega)$, shows a
characteristic resonance peak as we sweep through the resonance.

\begin{figure}
\hbox{ \resizebox{6.0cm}{!} {
\includegraphics{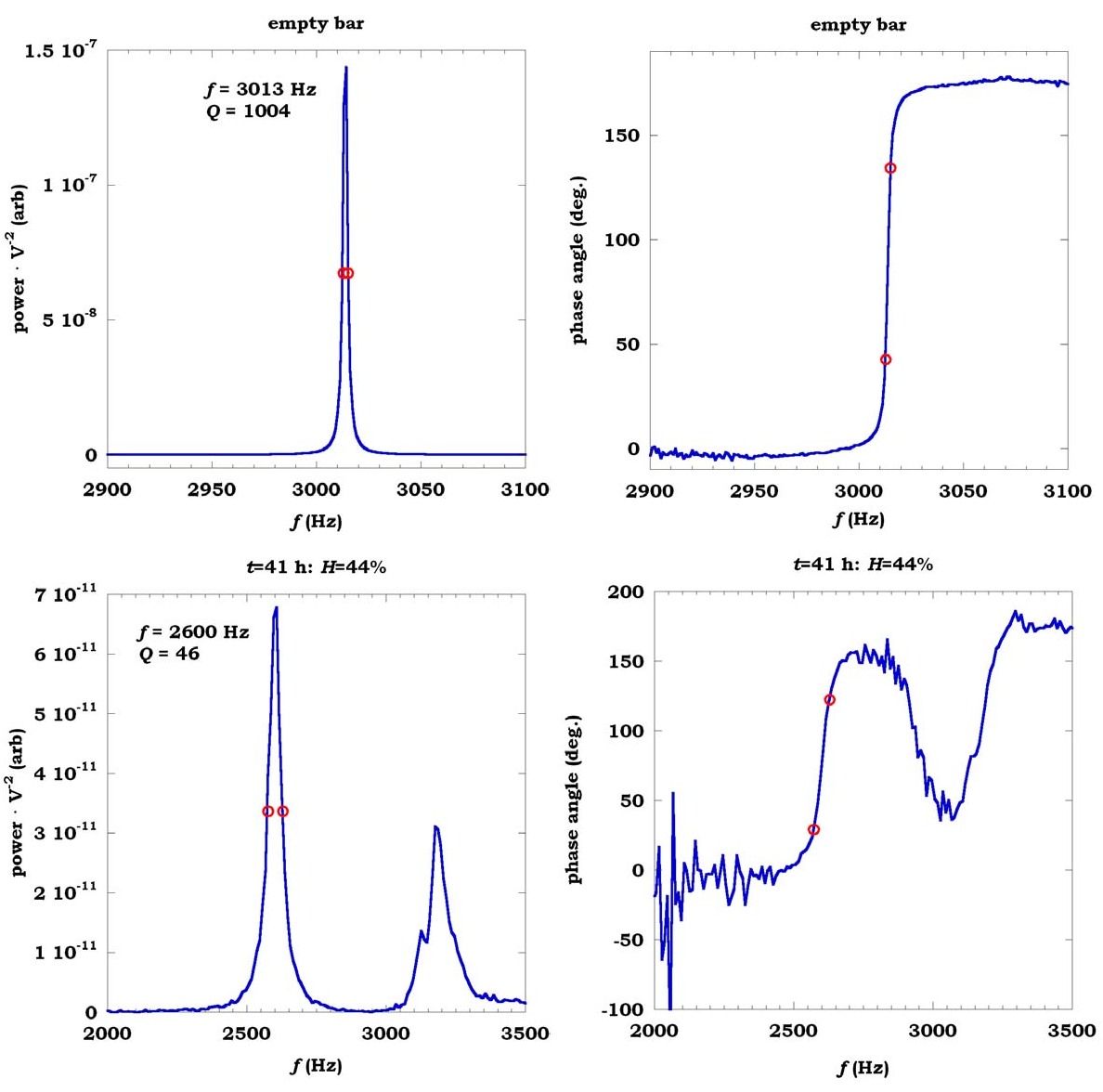}} } \caption {(Color online) Frequency
scan of resonance in bar without (top) and with (bottom) tungsten
granules.  The dissipated power is plotted in the left
column and the phase of the accelerometer, relative to the driving
voltage on the benders, is plotted in the right.  Note that, in this
example, there are two distinct resonances in the granules loaded
bar. The red circles mark the frequencies at which the power is
equal to half its peak value.  Note that there is a $90^o$ degree
phase shift between the two half-max points, as expected.}
\label{resonant_bar}
\end{figure}

We take the power dissipated in the vibrating bar to be
$\mathcal{P}(\omega) \propto \; \mid a(\omega) \mid ^2/\omega^2$.
This quantity is plotted, in arbitrary units, in Figure
\ref{resonant_bar}.  The real part of the resonance frequency,
$f_R$, and the quality factor, Q, may be taken from the peak and the
full-width at half maximum of this curve.  See pages 23 {\it ff.} of
Kinsler and Frey \cite{Kinsler}.

This technique of finding $f_R$ and $Q$ works well when there is an
isolated peak and the background power absorption is small. However,
in many cases during our experiments there are two nearby peaks with
significant overlap between them and/or there is a significantly
large background absorption. For such situations there are
established procedures for extracting $f_R$ and Q.  They assume
specific functional forms for the response function in which the
parameters therein are adjusted to achieve a best fit to the
response data \cite{mehl}.  We have taken a different approach which
assumes only that the data represents a sampling on the real axis of
an analytic function of complex frequency.

We record the complex-valued data for the acceleration of the bar,
$a(\omega_i)$, as we do for the cup, where $\{\omega_i\}$ is the set
of discrete measurement frequencies.  We analytically continue the
auxiliary function $g\stackrel{def}{=}1/a$ using a rational function
technique (Bulirsch-Stoer algorithm) \cite{numrec} (See also
Appendix B).  It is relatively simple to search for a zero of this
analytically continued function [$g(\omega_R) = 0$] using Muller's
method \cite{numrec}.   We have \beq{Q2} \omega_R = 2 \pi f_R[1-
i/(2Q)] = 2 \pi f_R -i\alpha \:\:\:,\eeq where $\alpha$ is the decay
rate of the mode.  We find that this technique is highly reliable.
As far as we are aware, it has not previously been reported in the
literature.

We see from Figure \ref{resonant_bar} that there is a very
significant effect due to the addition of the loose grains.  When
the cavity is empty, and the frequency scanned with 0.5 (Hz)
increments, we find that, in the range $100 {\rm (Hz)} < f < 5 {\rm (kHz)}$, there is only a single resonance: $f_0({\rm empty\:bar}) = 3014\:{\rm (Hz)} \pm
0.1\%$ and ${\rm Q}_0(\rm{exp}) = 1004\pm 15\%$.  When the cavity is
filled with the tungsten particles, and the bar is scanned in 10 (Hz)
frequency increments, there is a significant frequency shift for the
main resonance, $f_R ({\rm loaded})= 2600 \:{\rm (Hz)}$, and increase
in attenuation: ${\rm Q}_R(\rm{loaded}) = 46$. It is this latter
feature that makes granular dampening an attractive possibility for
reducing unwanted structure-borne sound.  Moreover, there is a new,
additional mode that is not present in the unloaded bar.  The major
thrust of this article is that these resonant bar characteristics
can be computed directly, based on the measured effective mass,
$\tilde{M}(\omega)$; a detailed comparison between theoretical
predictions and experimental results, as a function of humidity, is
presented later in Section \ref{sec:granmdehum} using a theory
developed in Appendix A.

\subsection{Sample Handling}\label{sec:samphand}

\begin{figure}
\hbox{ \resizebox{8.5cm}{!} {
\includegraphics{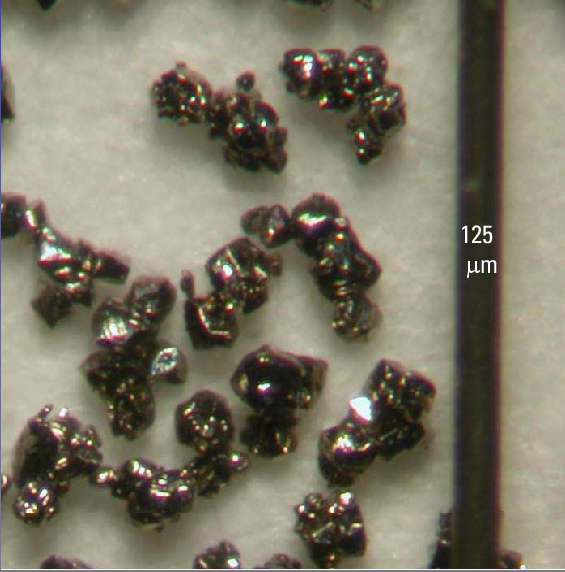}} }\caption{(Color online) Photomicrograph of some tungsten granules used in the
experiment.  There is a 125 $\mu$m wire at the right, for
comparison.} \label{W_granules_photo}
\end{figure}

A major concern for us is that we need to be able to prepare the
loose grains in the two cavities in a reproducible manner.  This
will allow us to make meaningful predictions about the resonant
properties of the bar, using the measured effective mass in the
shaker cup.  Granular media are notoriously hard to prepare
reproducibly because they often configure themselves into
meta-stable states.  For spherical grains it is relatively easy to
get the grains into a state approximating random close packing,
whose properties are quite reproducible \cite{sphere_pack}.  Our
granular systems, however, do not lend themselves easily to this.
This is because each granule consists of several sand-grain like
particles fused together.  A photomicrograph of a few of these
granules is shown in Figure \ref{W_granules_photo}.  Although the
main features of the effective mass are certainly reproducible, the
details of the structure in $\tilde{M}(\omega)$ can vary
significantly enough from one filling to the next that, unless
precautions are taken, it prevents an accurate theory-experiment
comparison in the bar.

\begin{figure}
\hbox{ \resizebox{8.5cm}{!} {
\includegraphics{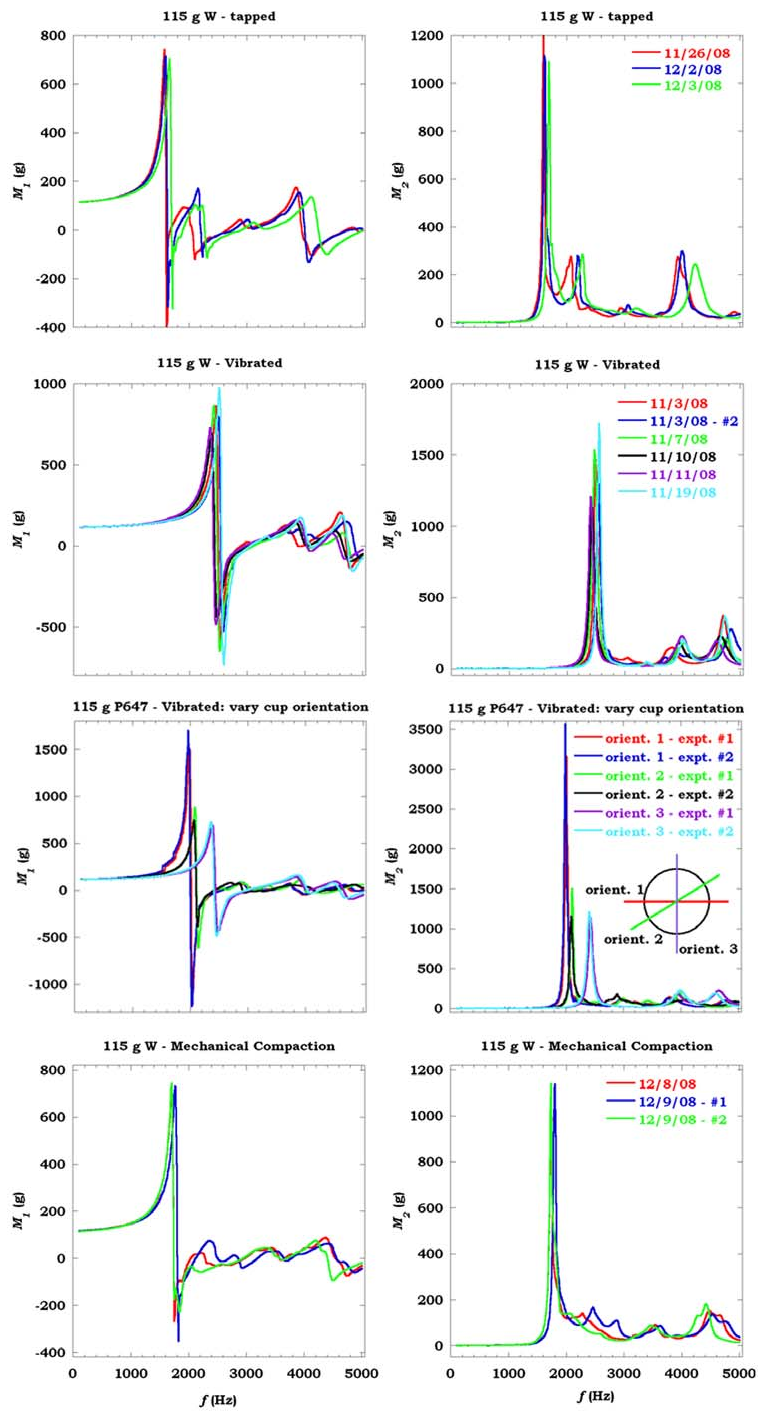}} }\caption{(Color online) Effective mass,
$\tilde{M}(\omega)$, measured on subsequent packings of the shaker
cup using four different loading protocols.  First Row: simple
filling and subjective tapping.  Second Row: Sinusoidal vibration,
aligned cup. Third Row: Sinusoidal vibration, different cup
orientations. Fourth Row: Mechanical compaction with commercial
press.} \label{repro}
\end{figure}

In Figure \ref{repro} we show the reproducibility of the effective
mass for consecutive runs using each of the different protocols.  In
the first row we show the results of filling while occasionally
tapping to settle the grains.  Next, we show the result of
vibrating the particulate medium at a frequency of 1 (kHz) over a
range of acceleration amplitudes, $0 < a
 < 30 \;(m\cdot s^{-2})$ with a 93 g stainless steel plug resting on the free
surface. The plug ensures that the grains at the surface experience
a static pressure similar to that experienced at the bottom of the
particle pack. The protocol consists of systematically increasing
the acceleration amplitude in steps of 5 $(m\cdot s^{-2})$, holding for 2 min
at each amplitude before increasing to the next level. After
reaching the maximum acceleration amplitude [$30 \;(m\cdot s^{-2})$] the
procedure is reversed.  Therefore, the sample is exposed to 6
different acceleration amplitudes, and the protocol consists of 11
total steps.  The procedure is similar to that employed by Nagel
{\it et al.} \cite{sphere_pack}.

The results indicate that vibrating the sample with a mass on the
free surface yields a fairly consistent $\tilde{M}(\omega)$. All of
the samples exhibit the same features, and these features are
observed at relatively consistent frequencies. The main resonance in
the grains is observed to occur over a slightly broader frequency
band (2.4 - 2.6 (kHz)) than that observed when the sample is tapped.
On the other hand, compared to that observed after the sample is
tapped there appears to be slightly better reproducibility in the
shape and position of the secondary features. It is interesting to
note that the higher degree of compaction achieved with the
vibration protocol shifts the main resonance to higher frequency, as
compared to that observed with the tapping protocol. In Section
\ref{sec:granmed} we will see that this shift can be interpreted as
an increase in effective sound speed in the granular medium.

Unfortunately, while reproducible, the effective mass so obtained
with this protocol varies with the orientation of the cup relative
to the shaker on which it is mounted.  This effect is demonstrated
in the third row of Figure \ref{repro}. The effective mass is
reproducible for a single orientation, but it varies with
orientation. Evidently the wobble effect we discussed earlier is big
enough to effect the packing of the grains, when we vibrate at these
high amplitudes. Therefore, the vibration protocol is not optimal
for handling the samples prior to a transferability experiment
because we cannot insure that we can duplicate the motion of the bar
and the cup as we vibrate it at high amplitudes during the
preparation phase.

The most successful method for developing a reproducible loading was
to mechanically compact the grains after they had been loaded in
their respective cavities.  This protocol consists of using a
mechanical testing instrument to impose a sinusoidal stress on the
free surface.  To promote a uniform imposition of the stress over
the free surface of the grains, we use a stainless steel plunger
with a rubber pad glued to the bottom surface.  First, a static
stress of 59.2 (kPa) is imposed on the granular medium. Then, a
sinusoidal stress is imposed on the system consisting of 200 cycles
at a frequency of 0.25 (Hz). The stress amplitude is systematically
varied between 39.5 (kPa) and 118.5 (kPa), in steps of 39.5 (kPa). To
prevent unloading the system, the static stress is increased by 39.5
(kPa) for each equivalent change in stress amplitude. After the
maximum stress amplitude is achieved, the procedure is repeated in
reverse. So the first two steps consist of increasing, and the
second two steps consist of decreasing the stress amplitude.  We
have limited the maximum stress on the grains to the low value of
118.5 (kPa) [= 1.185 (bar)] specifically in order to ensure that we are
not physically damaging any of the grains.  Overall the system is
exposed to 1000 stress cycles (5 sets of 200) at systematically
varied stress amplitudes.  In order to test the reproducibility of
this protocol we have repeated the measurement by dumping out the
grains, repacking the same amount by weight of fresh grains using
the compaction technique, measuring $\tilde{M}(\omega)$, dumping out
the grains, repacking, etc.  For the purposes of comparing theory
vs. experiment in the bar data [Section \ref{sec:granmdehum}], we
used the mechanical compaction protocol.

\section{Dampening of Structure-Borne Sound and
the Effective Mass of Granular Media}\label{sec:perth}

Subject to the validity of a few simple assumptions, it is possible
to use the measured effective mass, $\tilde{M}(\omega)$, to compute
its dampening effect on an elastic structure having an ``identical"
grain-filled cavity.  Let us suppose that there is an acoustic
structure which has a resonant frequency $\omega_0$ when the cavity
is empty. The resonant frequency becomes complex-valued in the
presence of the granular dampening mechanism: $\omega = \omega_0
+\Delta \omega$. Here, ${\rm Imag}(\omega)$ describes the ring-down
rate of the mode in the time domain or, equivalently, the quality
factor of the resonance in the frequency domain: Q $= - {\rm Re}
(\omega)/[2 \:{\rm Im}(\omega)]
>0$.  We make the approximation that the grains contribute an
additional mass-loading which is localized at position ${\bf x_1}$.
Thus, the density may be written as \beq{density} \rho({\bf x}) =
\rho_0({\bf x}) + \tilde{M}(\omega) \delta({\bf x - x_1}) \eeq where
$\rho_0({\bf x})$ is the point-by-point density of the structure
when the cavity is empty.  This assumption is essentially a
statement that we are considering only those normal modes of the
structure whose wavelengths are much larger than the dimensions of
the cavity.  We make the further assumption that the grain-filled
cavity contributes negligibly to any change in the effective elastic
moduli of the host structure. i.e.  The effective moduli of the
grains, as seen by the walls of the cavity, are much smaller than
those of the host material. In Section \ref{sec:granmed} we show
that this is an excellent approximation for the effective masses and
structures we are considering here.

The equations of elasto-dynamics may now be recomputed using this
modified, and now frequency dependent and complex-valued, density.
 We demonstrate how to do this sort of computation for the specific case of the
fundamental flex mode in a rectangular bar in Appendix A.

It is most useful, however, to consider the results of the much
simpler perturbation theory, which is valid to first order in
$\tilde{M}(\omega)$, considered as a small perturbation. The
displacement field obeys the usual equations of motion: \beq{EoM}
-\rho \lambda u_i = [C_{ijkl}u_{k,l}]_{,j} \:\:\:.\eeq Here, $\rho$
and $C_{ijkl}$ are the position dependent density and elastic
constants of the material, $u_i({\bf x})$ is the $i$-th component of
the displacement field, a comma denotes differentiation w.r.t that
coordinate, and summation over repeated indices is understood;
$\lambda \equiv \omega^2$ is the eigenvalue for the problem. Written
in this manner, Equation \rf{EoM} applies to any position-dependent
material constants, $\rho({\bf x})$ and $C_{ijkl}({\bf x})$,
including those with step discontinuities.  We consider resonances
such that either the displacement field vanishes on the boundary
surface of the structure, ${\bf u} \mid_S \equiv 0$, or the stress
tensor vanishes on that surface, $C_{ijkl}u_{k,l} \mid_S \equiv 0$.
Moreover, we assume that the elastic constants, $C_{ijkl}$, and the
unperturbed density, $\rho_0$, are real-valued, thus guaranteeing
that $\omega_0$, the resonance frequency when there are no grains in
the cavity, is also real-valued.

Now, if the density is perturbed to take on a new value, $\rho =
\rho_0 + \Delta \rho$ then there is a corresponding change in both
the eigenvalue $\lambda = \lambda_0 + \Delta \lambda$, and in the
eigenvector, ${\bf u} = {\bf u}^0 + \Delta {\bf u}$.  Substituting
into Eq. \rf{EoM} and collecting all the first order changes one has
\beq{1storder} -\Delta \lambda \rho_0 u^0_i - \lambda_0 \rho_0
\Delta u_i -\lambda_0 \Delta \rho u^0_i = [C_{ijkl} \Delta
u_{k,l}]_{,j}\:\:\:.\eeq  Multiply Eq. \rf{1storder} by $[u^0_i]^*$
(sum on $i$) and integrate over all space.  The term
 $\int [u^0_i]^*[C_{ijkl} \Delta u_{k,l}]_{,j}\: dV$ can be integrated
by parts; the surface term vanishes because of the assumed boundary
condition (above). The remaining volume integral cancels the term
$-\lambda_0 \int \rho_0 [u^0_i]^* \Delta u_i \: dV$ (because ${\bf
u}_0$ satisfies the zero-order equation).  Thus, \beq{dellam}\Delta
\lambda= -\frac{\lambda_0 \int \Delta \rho {(\bf x}) \mid {\bf u^0
(x)} \mid ^2\: dV}{\int \rho_0({\bf x})\mid {\bf u^0 (x)} \mid
^2\:dV} \:\:\:.\eeq  Applied to the case of interest, Eq.
\rf{density} for which $\Delta \rho({\bf x}) = \tilde{M}(\omega)
\delta({\bf x - x_1})$, this now reads: \beq{pertheory} \Delta
\omega = - \left [\frac{\omega_0  I_1}{2 M_0}\right ] \tilde
{M}(\omega_0)\:\:\:.\eeq where \beq{Ione} I_1 = \frac {M_0 \mid {\bf
u^0 (x_1)} \mid ^2}{\int \rho_0({\bf x})\mid {\bf u^0 (x)} \mid
^2\:dV} \:\:\:,\eeq in which ${\bf u^0 (x)}$ is the displacement
field, and $\rho_0({\bf x})$ is the density, when the cavity is
empty of grains.  $M_0 = \int \rho_0({\bf x})\:dV$ is the total mass
of the structure when there are no grains in the cavity. Written in
this manner, $I_1$ is dimensionless. In order for this sort of
theory to be valid, the granular media in the two cavities must be
substantially the same. In Section \ref{sec:granmdehum} we make such
a theory-experiment comparison, but we have had to go beyond the
simple perturbation theory result for two reasons: (1) The effective
mass, $\tilde{M}(\omega)$, is not always small.  It can, in fact,
take on values larger than that of the steel bar. Perturbation
theory cannot give an accurate description in these cases. (2) Some
of the modes seen in the bar + grains system are basically modes
within the grains [i.e. poles of $\tilde{M}(\omega)$] modified by
the effect of the bar. Perturbation theory is silent as to the
properties of these modes.  Nonetheless, for the modes which are
primarily bar-like, Equation \rf{pertheory} gives a useful intuitive
way to think about the effects of the granular medium on the
resonance frequency and Q. Specifically, $M_1(\omega_0)$ determines
the shift in the resonance frequency and $M_2(\omega_0)$ determines
the lowered Q, as is clear from Equation \rf{pertheory}.

We note in passing that the effective mass is, in reality, a tensor,
viz $\tilde{M}_{ij}(\omega)$, reflecting the fact that gravity plays
a major role in establishing stiffness and dampening at the
contacts.  Equation \rf{EoM} has an obvious generalization to the
case of a tensorial density and Equation \rf{pertheory} becomes
\beq{tensor} \Delta \omega = -\frac{\omega_0 u_{i}^{0*} (x_1)\tilde
{M}_{ij}(\omega_0) u^0_{j} (x_1)} {2\int \rho_0({\bf x})\mid {\bf
u^0 (x)} \mid ^2\:dV} \:\:\:.\eeq  In this article we are
considering only situations in which the cavity motion is strictly
along the z-axis, parallel to the force of gravity.  Thus the only
relevant component of the effective mass being considered here is
$\tilde {M}_{zz}(\omega)$ which we shall henceforth denote without
the subscripts, but with this understanding that there are other
nonzero components of the tensor.

\section{Properties of $\tilde{M}(\omega)$}\label{sec:genprop}
In this Section we first discuss some general properties of
$\tilde{M}(\omega)$ considered as a causal response function.  Next,
we investigate some properties of the system in which we idealize
the grains as being rigid particles which interact with their
neighbors via contact forces, of which there may be different kinds.
A specific motivation here is to analyze the high frequency behavior
seen in our own measurements of $\tilde{M}(\omega)$ within this
discrete particles context.  Finally,  some of the features we
observe in $\tilde{M}(\omega)$ are suggestive of a collective motion
in which the displacement varies slowly from grain to grain.  This
suggests the possible approximate validity of continuum models, two
of which we present here.

\subsection{General}\label{sec:genpropA}
The effective mass, $\tilde{M}(\omega)$, is a causal response
function in that one may, in principle, apply an arbitrary
time-based protocol to the acceleration of the cup and measure the
force induced by this protocol.  As such it has several general
properties, which we summarize here.  These properties follow on
general principles as described in e.g. Landau and Lifshitz
\cite{landau}.  $\tilde{M}(\omega)$ is the fourier transform of a
real-valued memory function: \beq{memory} \tilde{M}(\omega) =
\int_0^\infty \chi(t) e^{i \omega \tau} \:d\tau\:\:\:. \eeq
Causality considerations specifically restrict the range of
integration to positive values of $\tau$ only.  That is, in the time
domain, the force and the acceleration are related to each other via
the memory function, $\chi(t)$: \beq{timedomain} F(t) =
\int_0^{\infty} \chi(\tau) a(t-\tau)\;d\tau \:\:\:.\eeq Inasmuch as
only the past history of the acceleration matters in Equation
\rf{timedomain} its fourier transform leads to Equation \rf{memory}.
If the frequency, $\omega$, is extended to take on complex values,
we see that $\tilde{M}(\omega)$ is regular everywhere in the
upper-half complex plane.  We also see from Eq. \rf{memory} that
\beq{symmetry} \tilde{M}(-\omega*) = \tilde{M}^*(\omega) \:\:\:,\eeq
where an asterisk signifies complex conjugation.  These
considerations lead immediately to the usual Kramers-Kronig
relations between the real and the imaginary parts: \beq{KK1}
M_1(\omega) = \frac{2}{\pi} P\left [\int_0^\infty
\frac{xM_2(x)}{x^2-\omega^2} dx \right ]\eeq \beq{KK2} M_2(\omega) =
-\frac{2 \omega}{\pi} P \left [\int_0^\infty
\frac{M_1(x)}{x^2-\omega^2} dx \right ]\:\:\:, \eeq where $\omega$
and $x$ take on real values only and $P[\:]$ denotes principal part
of the argument.  (We are assuming there are no singularities on the
real axis and that $\lim_{\omega \rightarrow \infty}
\tilde{M}(\omega) = 0$. See Section \ref{sec:genprop} {\bf B},
below.)

One may just as well consider a time-based protocol for an applied
force and measure the induced acceleration of the cup.  Therefore
$1/\tilde{M}(\omega)$ is also a causal response function and all the
results quoted above apply to it as well, except for the
Kramers-Kronig relations. Thus, $\tilde{M}(\omega)$ has no zeroes,
poles, branch points, or singularities of any kind anywhere in the
upper half plane.  Inasmuch as only the constant functions are
analytic and bounded everywhere in the complex plane,
$\tilde{M}(\omega)$ must have singularities of some sort in the
lower-half plane.

The power dissipated in the system, averaged over one cycle of
oscillation, is \beq{power} \begin{array}{rcl} \mathcal{P} &=&
\langle Re[F] Re[v]
\rangle\\
&&\\
& = & (1/2)Re[F\;v^*]\\
&&\\
&=& (1/2) \; \omega M_2(\omega) \mid v \mid^2\:\:\:,
\end{array}
\eeq where $v=i a/\omega$ is the velocity of the cup. Since
$\mathcal{P}$ is always non-negative, one has $\omega M_2(\omega)
\ge 0$, for real values of the frequency.  We see that this property
is borne out by the data in Figures \ref{F_a_m} and \ref{repro}.

\subsection{Systems of Discrete Particles}
Let us consider a model in which each grain is considered to be
rigid except for the region near the contacts with its neighboring
particles.  Let ${\bf X}_i$ be the equilibrium position of the
center of mass of the i-th particle, whose mass is $m_i$, and ${\bf
u}_i$ be its displacement from equilibrium.  Similarly,
${\bftheta}_i$ is the librational angle of rotation.  If two
neighboring particles translate or rotate such that their points of
contact would move relative to each other there will be a restoring
force due to the contact forces.  The linearized equation of motion
for the i-th particle is \beq{EoM2} \begin{array}{c}-m_i \omega^2
{\bf u}_i = - {\bf K}_{iw} \cdot [{\bf u}_i + {\bftheta}_i \times
{\bf d}_{iw} - {\bf W}] \\
\\
+ \sum_j {\bf K}_{ij} \cdot [{\bf u}_j +{\bftheta}_j \times {\bf
d}_{ji} - {\bf u}_i - {\bftheta}_i \times {\bf d}_{ij}]\\
\end{array}\;\;\;,\eeq where $ {\bf d}_{ij}$ is the vector from ${\bf X}_i$ to the
point of contact with the j-th grain.  It is understood that the
tensor ${\bf K}_{ij}(\equiv {\bf K}_{ji})$ is nonzero only for
grains actually in contact with each other.  (We assume there is at
most one contact per pair.)  Similarly, $ {\bf d}_{iw}$ and ${\bf
K}_{iw}$ refer to grains that are in contact with the surfaces of
the cavity, whose rigid displacement is ${\bf W}$.

The equation of motion for the angular variables is
\beq{EoM3}\begin{array}{c} -\omega^2 {\bf I}_i \cdot {\bftheta}_i =
- {\bf d}_{iw} \times {\bf K}_{iw} \cdot [{\bf u}_i +
{\bftheta}_i \times {\bf d}_{iw} - {\bf W}]\\
\\
 + \sum_j {\bf d}_{ij} \times {\bf K}_{ij} \cdot [{\bf u}_j
+{\bftheta}_j \times {\bf d}_{ji} - {\bf u}_i - {\bftheta}_i \times
{\bf d}_{ij}] \end{array}\;\;\;,\eeq where ${\bf I}_i$ is the moment
of inertia tensor for the i-th particle.

In the special case that the particles are identical spheres we have
${\bf d}_{ij} = (1/2)[{\bf X}_j -  {\bf X}_i]$ and the spring
constant tensor may be written in terms of normal (N) and transverse
(T) stiffnesses as \beq{sphpart} {\bf K}_{ij} = k_{ij}^N \hat{{\bf
d}}_{ij}\hat{{\bf d}}_{ij} + k_{ij}^T[{\bf I} - \hat{{\bf
d}}_{ij}\hat{{\bf d}}_{ij}] \;\;\;,\eeq where $\hat{{\bf d}}_{ij}$
is the unit vector and we use dyadic notation.  Similarly for the
contacts with the walls.  An example here would be Hertz-Mindlin
contact forces in which the stiffnesses increase with increasing
static compression but we also consider forces of adhesion,
capillarity, etc.

It is understood that, generally, each of the elements of the
tensors $ {\bf K}_{ij}$ or ${\bf K}_{iw}$ are complex-valued and
frequency dependent reflecting the microscopic origin of the
dissipation.  For example, one may take \beq{fdep}{\bf
K}_{ij}(\omega) = {\bf K}_{ij}^0 - i \omega {\bf B}_{ij}\eeq in
which the second term describes an inter-particle force proportional
to the difference in velocity of the two grains.  The tensor $ {\bf
B_{ij}} $ is analogous to a dampening parameter.  In general Eq.
\rf{fdep} represents simply the first two terms in the Taylor's
series expansion of ${\bf K}(\omega)$.  These ``springs" may have
rheological properties of their own.  For example, if there is an
internal degree of freedom with relaxation time $\tau$ it is easy to
show that \beq{relaxtime} {\bf K}_{ij}(\omega) = {\bf
K}_{ij}^{\infty} + \frac{{\bf K}_{ij}^0 - {\bf
K}_{ij}^{\infty}}{1-i\omega \tau}\;\;\;.\eeq The derivation of Eq.
\rf{relaxtime} parallels that of Eq. (78.6) in Landau and Lifshitz
\cite{llfm}.  A Taylor's series expansion of Eq. \rf{relaxtime} for
small $\omega$ has the form of Eq. \rf{fdep} for the first two
terms.

Notwithstanding the foregoing remarks there are some general
conclusions one can draw from Eqs. \rf{EoM2} and \rf{EoM3}.  First,
the total force which the cavity exerts on the grains is
\beq{netforce}
\begin{array}{rcl} {\bf F} &=&  - \sum_i {\bf K}_{iw} \cdot [{\bf u}_i +
{\bftheta}_i \times {\bf d}_{iw} - {\bf W}]\\
&&\\
& = &-\omega^2 \sum_i m_i {\bf u}_i\;\;\;,
\end{array}\eeq where the second equality follows because the
interparticle forces cancel, by Newton's third law, as is clear from
Eq. \rf{EoM2}.  Secondly, one can formally write the effective mass
in terms of the normal-mode frequencies of the system: \beq{normal}
\tilde{{\bf M}}(\omega) = \sum_n \frac{{\bf A}_n}{\omega-\omega_n}
\eeq where $\omega_n$ are the complex-valued frequencies for which
Eqs. \rf{EoM2} and \rf{EoM3} have nontrivial solutions when ${\bf
W}$ is set equal to zero.  Each matrix ${\bf A}_n$ represents the
strength of each resonance.  Also, from Eqs. \rf{EoM2} and \rf{EoM3}
it is clear that when the frequency tends to zero one has
$\lim_{\omega \rightarrow 0}{\bf u}_i = {\bf W}$ and $\lim_{\omega
\rightarrow 0}{\bftheta}_i = 0 $. Therefore, in this limit one has,
from the second of Eq. \rf{netforce}, ${\bf F} = -\omega^2 M_0 {\bf
W}$ where $M_0 = \sum_i m_i$ is the total static mass of the grains.
Therefore, from the definition of the effective mass, one has:
\beq{zero} \lim_{\omega\rightarrow 0}\tilde{{\bf M}}(\omega) = M_0
{\bf I}\;\;\;, \eeq which seems obvious.

The high frequency limit of the effective mass also has a simple
form.  From Eqs. \rf{EoM2} and \rf{EoM3} one has $\lim_{\omega
\rightarrow \infty}{\bf u}_i = 0$ and $\lim_{\omega \rightarrow
\infty}{\bftheta}_i = 0 $, because of the dominance of inertial
effects.  i.e.  The particles do not move at all, in this limit.
Therefore, in this limit the first of Eq. \rf{netforce} implies
${\bf F} = \sum_i \lim_{\omega \rightarrow \infty} {\bf
K}_{iw}(\omega)\cdot {\bf W}$ which, in turn, implies \beq{hflim}
\lim_{\omega \rightarrow \infty} \tilde{\bf M}(\omega) = -
\frac{\lim_{\omega \rightarrow \infty}\sum_i {\bf
K}_{iw}(\omega)}{\omega^2} \:\:\:.\eeq  The high frequency behavior
of $\tilde{\bf M}(\omega)$ is controlled by the behavior of those
``springs" connecting the particles directly to the cavity motion.
If, for example, these are strictly dampened springs, such as
implied by Eq. \rf{fdep}, then \beq{hflim2} \lim_{\omega \rightarrow
\infty} \tilde{\bf M}(\omega) = i\frac{\sum_i {\bf B}_{iw}}{\omega}
\:\:\:,\eeq which is predominantly imaginary valued in this limit.
On the other hand, if the individual springs have their own
rheological behavior, such as that implied by Eq. \rf{relaxtime},
then \beq{hflim3} \lim_{\omega \rightarrow \infty} \tilde{\bf
M}(\omega) = -\frac{\sum_i {\bf K}_{iw}^{\infty}}{\omega^2}
\:\:\:.\eeq  Should this latter case hold then comparison with Eq.
\rf{KK1} immediately gives an f-sum rule on the absorptive part of
the effective mass: \beq{fsum} \int_0^{\infty} \omega \:{\rm
Imag}[{\tilde{\bf M}}(\omega)]\;d\omega = \frac{\pi}{2} \sum_i {\bf
K}_{iw}^{\infty}\:\:\:.\eeq  So if those grains that are in contact
with the walls of the cavity have spring constants which are
real-valued in the high frequency limit then the integrated
absorptive part of the effective mass is determined solely by those
high frequency spring constants, independent of the details of the
grain-grain contact forces.

The potential usefulness of the results of this subsection is that
the high frequency limit of the measured $\tilde{M}(\omega)$ may be
compared against Eqs. \rf{hflim} and its variants.  In our own
experiments we are not able to probe frequencies high enough to
observe this behavior, as discussed in Section \ref{sec:granmed},
below.

\subsection{Continuum Models}\label{sec:contmod}
 On a semi-quantitative level, we may
understand the general features of our experimental results, such as
those in Figures \ref{F_a_m} and \ref{repro}, in terms of two
over-simplified continuum models whose main purpose will be to allow
us to deduce approximate parameter values from our data.\\

\begin{figure}
\hbox{ \resizebox{7.5cm}{!} {
\includegraphics{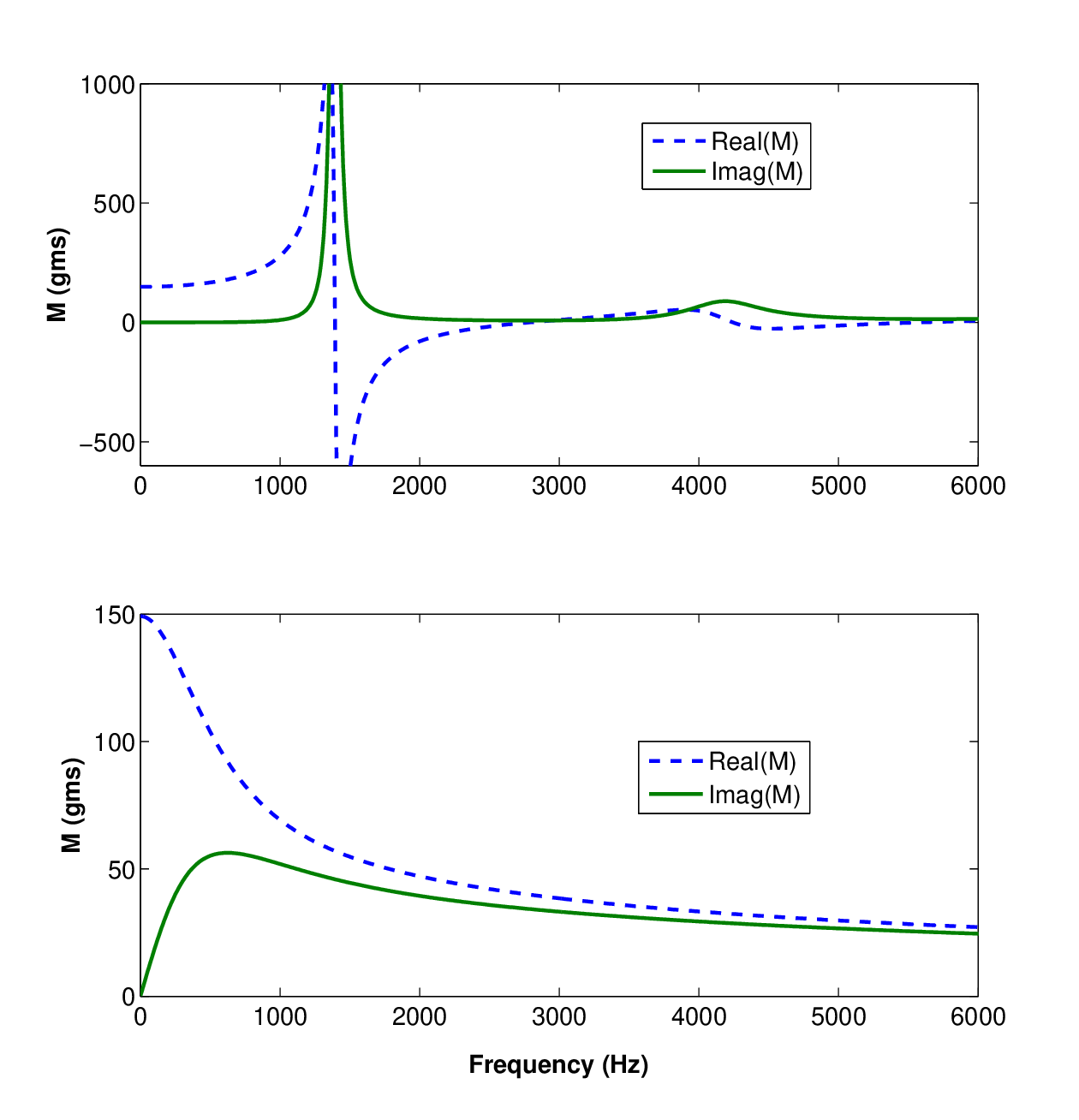}} } \caption {(Color online) (Top) Model I, a fluid with a lossy bulk modulus. (Bottom) Model II, a
highly compressible, highly viscous fluid. } \label{model}
\end{figure}

{\bf Model I:} The granular medium is considered to be a lossy
fluid, with
 negligible viscous effects at the walls; that is, the viscous skin depth \cite{landau} $\delta =
 \sqrt{2 \eta/(\rho \omega)}$ is small compared to the radius of the cup, $a$.  Here, $\eta$
is the viscosity of the fluid; $\rho$ is the density.  The effective
mass is simply \beq{fluid1}
 \tilde{M}(\omega) =  M_0 \tan(qL)/(qL)\:\:\:,
 \end{equation}
 where $L$ is the height of the fluid column, $q =  \omega \sqrt{\rho/K} $ is the  wave vector in the
 fluid in terms of a (lossy) bulk modulus,
 $K= K_0[1-i \omega \tau]$.  $M_0 = \pi
 a^2 L \rho$ is the static mass in the cup.  This model presupposes that
100\% of $\tilde{M}(\omega)$ is supported by the bottom surface of
the cup.  For ``small enough" values of the damping parameter,
$\tau$, resonance peaks, as seen in Figure \ref{model}a, occur when
$qL$ equals odd multiples of $\pi/2$ i.e. L equals odd multiples of
1/4 wavelength: $L = \lambda/4, 3\lambda/4, 5\lambda/4. \cdots$
These sharp resonances are examples of $f_g$, the resonances within
the granular medium/liquid.  The values of $K_0$ and $\tau$ are
chosen to mimic the observed frequency position and resonance width,
respectively, in the experiments. There is a second resonance in
Figure \ref{model}a around 4500 (Hz), but the width of that resonance
is nines times as large as the first one, so it is scarcely visible
in the plot.

We emphasize that this last feature, the resonance frequencies being
in the ratio 1:3:5:7 ..., is an artifact of three assumptions in
Model I:  (a) There is no shear rigidity or shear viscosity in the
material. (b) The sound speed is constant throughout the sample. (c)
The attenuation parameter, $\tau$, is small. We shall see
that all of these assumptions are violated, strictly speaking, under
the conditions of our experiments on real granular media.\\

 {\bf Model II:} The granular medium is considered to be a very viscous fluid, which is infinitely compressible.  This situation may be approximately
 correct if the filling depth of the cavity is much greater than the diameter
 so that most of $\tilde{M}(\omega)$
is borne by the walls of the cup, little by the bottom surface. By
solving the Navier-Stokes equation for oscillatory motion in a
cylindrical geometry \cite{landau} we find \beq{fluid2}
 \tilde{M}(\omega) = 2M_0 J_1(\kappa a)/[\kappa a J_0(\kappa a)].
 \end{equation}
Here $\kappa= \sqrt{i \rho \omega/\eta}$, and $J_k(z)$ is a Bessel
function of order $k$.  This model gives a broad peak with a slow
decay at higher frequencies as seen in Figure \ref{model}b, much
like that seen qualitatively in the granular data above the
fundamental resonance.  We note, for future use, that the high
frequency limit of Eq. \rf{fluid2} is \beq{hfluid} \tilde{M}(\omega)
\rightarrow 2 \pi a L \sqrt{i \eta \rho/\omega}\:\:\:. \eeq  This
limit holds when the viscous skin depth \cite{llfm}, $\delta =
\sqrt{2 \eta/\rho \omega}$ , is much smaller than the radius of the
cup. We suppose, then, that the result $\tilde{M}(\omega) \propto
\sqrt{i/\omega}$ holds generally, for other geometries, but with
different prefactors.

Is the high-frequency limit for this continuum model consistent with
what one might expect for a discrete system of particles, with
dampening due to their relative motion, namely Eq. \rf{hflim2}?
Though one may expect viscous-like dampening analogous to that
described by Eq. \rf{hfluid}, at high enough frequencies the viscous
skin depth becomes small compared to the inter-particle separation
and there is a crossover from $\tilde{M} \propto (1+i)\omega^{-1/2}$
to $\tilde{M} \propto i\omega^{-1}$. This can be demonstrated
explicitly if one considers a one-dimensional string of point
masses, separated a distance $b$ from each other, each of which
experiences a drag force proportional to the difference between its
velocity and its neighbor's, viz: $F=\gamma[v_j-v_i]$. This leads to
the equation of motion for the ensemble: \beq{1Dvd}-m\omega^2 u_j =
-i\omega \gamma(u_{j+1}-2u_j +u_{j-1})\:\:\: j=-N,...N \eeq subject
to the boundary condition \beq{1Dvda} u_{\pm(N+1)} = W \:\:\:.\eeq
Equation \rf{1Dvd} is a simple example of Equation \rf{fdep} in
which  $B_{ij} = \gamma$ if $i$ and $j$ are nearest neighbors and
$K_{ij} \equiv 0$. It is simple enough to solve for the effective
mass implied by this toy model. Let \beq{y1} y_{\pm}=\frac{2\gamma
-im \omega \pm i\sqrt{m^2 \omega^2 +4im \omega
\gamma}}{2\gamma}\:\:\:.\eeq The dynamic effective mass of one such
row is \beq{onerow} \tilde{M}_{1D}(\omega) =
\frac{2i\gamma}{\omega}\left [ \frac{y_+^{N+1} +y_-^{N+1}
-y_+^N-y_-^{N}}{y_+^{N+1} +y_-^{N+1}}\right ]\:\:\:.\eeq

\begin{figure}
\hbox{ \resizebox{8cm}{!} { \includegraphics{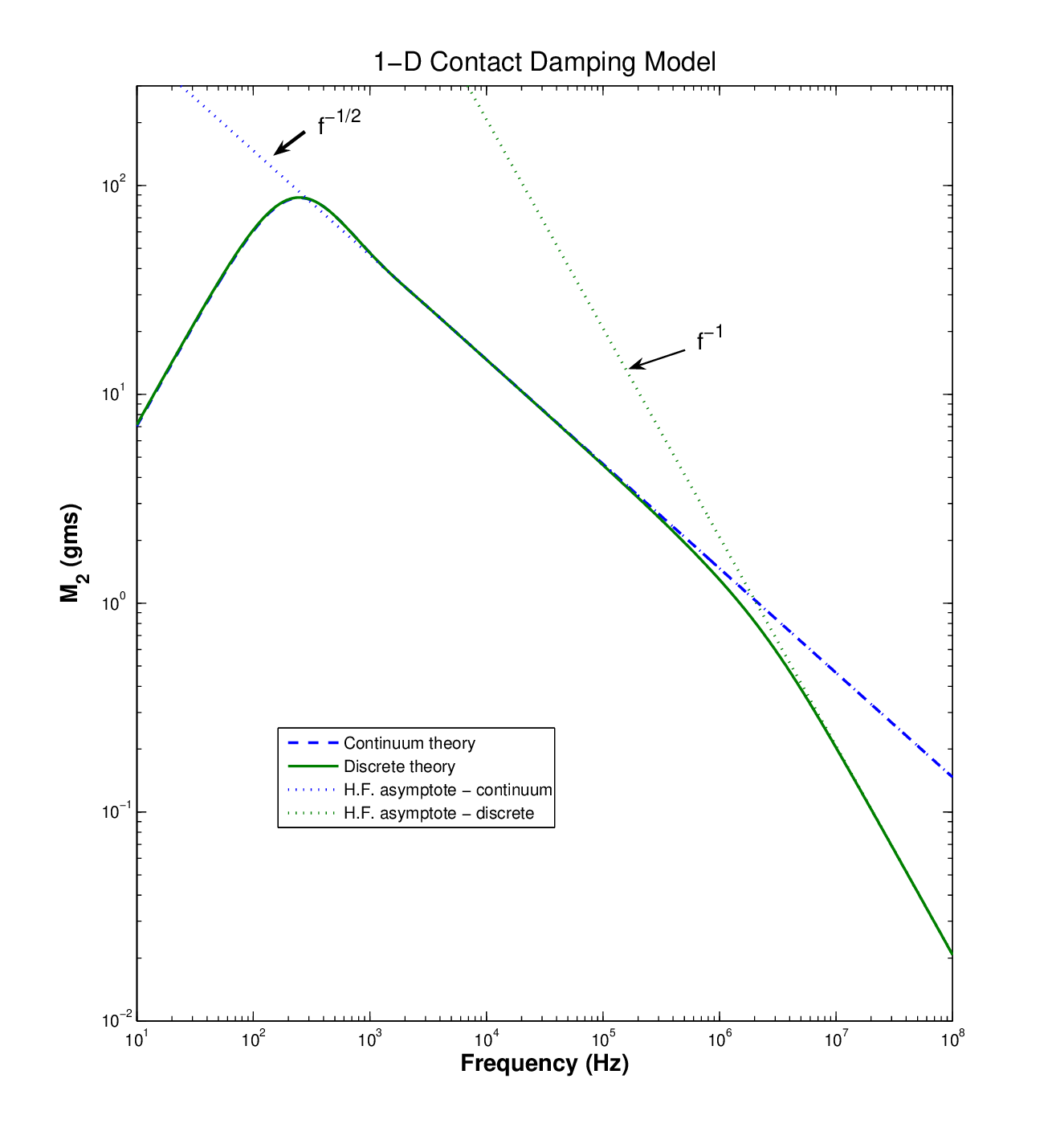}} }
\caption {(Color online) Comparison of a continuum versus a discrete
theory for a one-dimensional model of local contact dampening.  The
discrete model closely follows the continuum result except for
frequencies high enough such that the macroscopic viscous skin depth
becomes smaller than the interparticle separation.}
\label{one_D_contact}
\end{figure}

If we imagine that there is a sequence of these chains, in parallel
with each other, all connected to the walls then Eq. \rf{1Dvd} is a
discretized version of the linearized Navier-Stokes equation for
which the viscosity is $\eta \propto \gamma b$.  The effective mass
per unit area of the side wall implied by the continuum Navier
Stokes equation in this geometry is \beq{NS1D}
\tilde{M}_{NS}(\omega)/A =\frac{2\rho}{\kappa} \tan(\kappa
T/2)\:\:\:,\eeq where $T$ is the separation between the walls.
Equation \rf{NS1D} has a high-frequency limit analogous to Equation
\rf{hfluid}: \beq{hfNS1D} \tilde{M}_{NS}(\omega)/A \rightarrow 2
\sqrt{i \eta \rho/\omega}\:\:\:.\eeq  Taking into account the
appropriate normalization of the relevant constants one may directly
compare Eq. \rf{onerow} against Eq. \rf{NS1D}.  This is done in
Figure \ref{one_D_contact} where we plot the imaginary part of the
effective mass implied by each of the models.  They agree with each
other over much of the frequency range.  The peak around 300 (Hz)
occurs when the viscous skin depth is approximately equal to the
wall separation, $T$.  Above that frequency each model has a
frequency dependence $\tilde{M}_2 \propto \omega^{-1/2}$, as
expected from Eq. \rf{hfNS1D}.  When, however, the macroscopic
viscous skin depth, $\delta = \sqrt{2 \eta/\rho \omega}$, is
approximately equal to the inter-particle separation, $b$, the
discrete model, Eq. \rf{onerow}, crosses over to a behavior implied
by Eq. \rf{hflim2}: $\tilde{M}_2 \propto \omega^{-1}$. In Figure
\ref{one_D_contact} this crossover is visible around 3 (MHz).

For the purpose of the analysis of our data in Section
\ref{sec:granmdehum} we recapitulate the essential results of this
simple model.  If the frequency is high enough that the inter-grain
springs are dominated by the damping effect rather than the
stiffness, i.e. $\omega \gg K/B$ viz. Eq. \rf{fdep}, the system may
be described in terms of an effective viscosity.  If the frequency
is high enough that the viscous skin depth is small compared against
the dimensions of the cavity, one may expect $\tilde{M}_2 \propto
\omega^{-1/2}$.  For higher frequencies still, there may be a
crossover to $\tilde{M}_2 \propto \omega^{-1}$.

To be absolutely complete we point out that, just as contact damping
could lead to a crossover from $\tilde{M}_2 \propto \omega^{-1/2}$
to $\tilde{M}_2 \propto \omega^{-1}$ if the frequency is increased
high enough, it is also true that global damping could, in
principle, exhibit the exact opposite crossover behavior, from
$\tilde{M}_2 \propto \omega^{-1}$ to $\tilde{M}_2 \propto
\omega^{-1/2}$ if the frequency is raised high enough.  This is
because eventually the viscous skin depth in the surrounding air,
$\delta = \sqrt{2 \eta/\rho \omega}$, becomes small compared to
$\Lambda$, the connected throat size of the porous medium.  For air
at STP $\delta = 22 \mu$m at a frequency of $10^4$ (Hz), which is
still relatively large compared to the throat sizes of the pore
space in these materials, for which $\Lambda \approx 10 \mu$m.  See
References \cite{dynperm}.  The viscosity of air varies by less than
0.5 \% as the humidity changes from dry to fully saturated
\cite{turner}.  We do not expect to see this crossover in our
experiments.

To summarize this Section we may say that there are several distinct
possible origins to the frequency dependence of $\tilde{M}$.  Within
the context of the discrete model, such as embodied in Eqs.
\rf{EoM2} and \rf{EoM3}, one expects structure near those normal
mode frequencies which are relatively near the real axis, as implied
by Eq. \rf{normal}. To the extent that a continuum approximation may
be relevant for some of the lower lying modes, one may expect
structure when the (wavelength of sound/viscous skin depth) is
comparable to the cavity dimensions, examples of which appear in
Figures \ref{model}a, \ref{model}b, or \ref{one_D_contact}.  There
may be further structure at the frequencies for which the continuum
theory breaks down, an example of which is seen around 3 (MHz) in
Figure \ref{one_D_contact}.  And finally, there may be structure due
to the fact that the springs themselves exhibit a non-trivial
frequency dependence, an example of which is given by Eq.
\rf{relaxtime}.

\section{Effective Macroscopic Properties}\label{sec:macprop}
\subsection{Liquids}\label{sec:liquidEM}

\begin{figure}
\hbox{ \resizebox{7.5cm}{!} {
\includegraphics{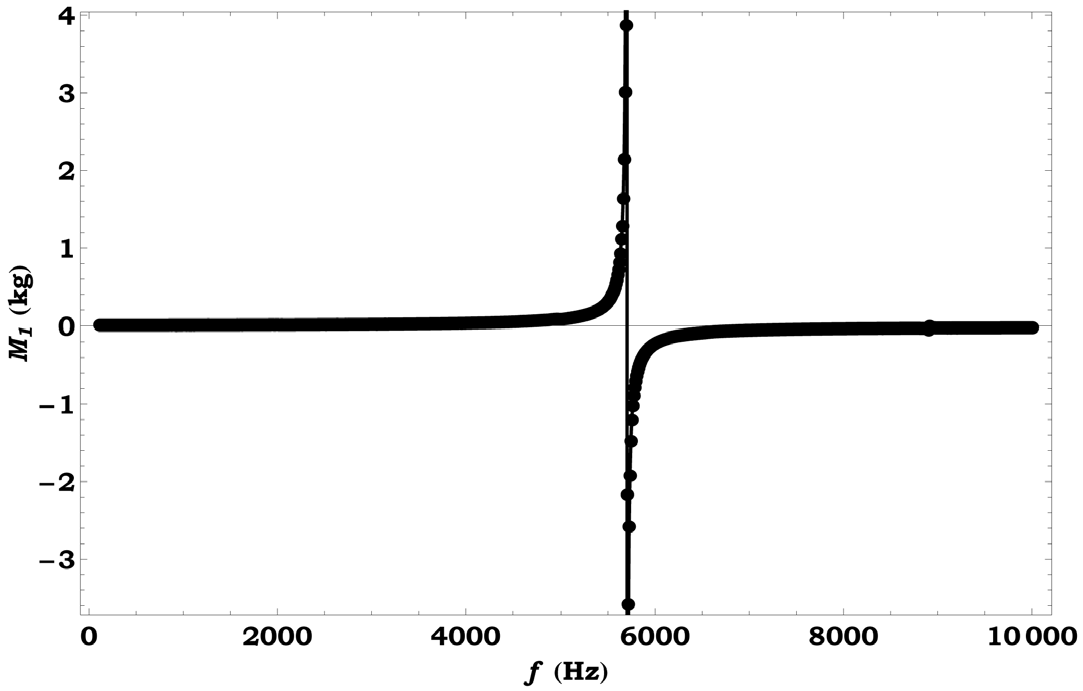}} } \caption {Real part of the
effective mass of a fluorocarbon fluid compared against the
theoretical prediction of Eq. \rf{fluid1}. The 1/4 wavelength
resonance is visible around 6 (kHz).} \label{fluorinert}
\end{figure}

\begin{figure}
\hbox{ \resizebox{7.0cm}{!} {
\includegraphics{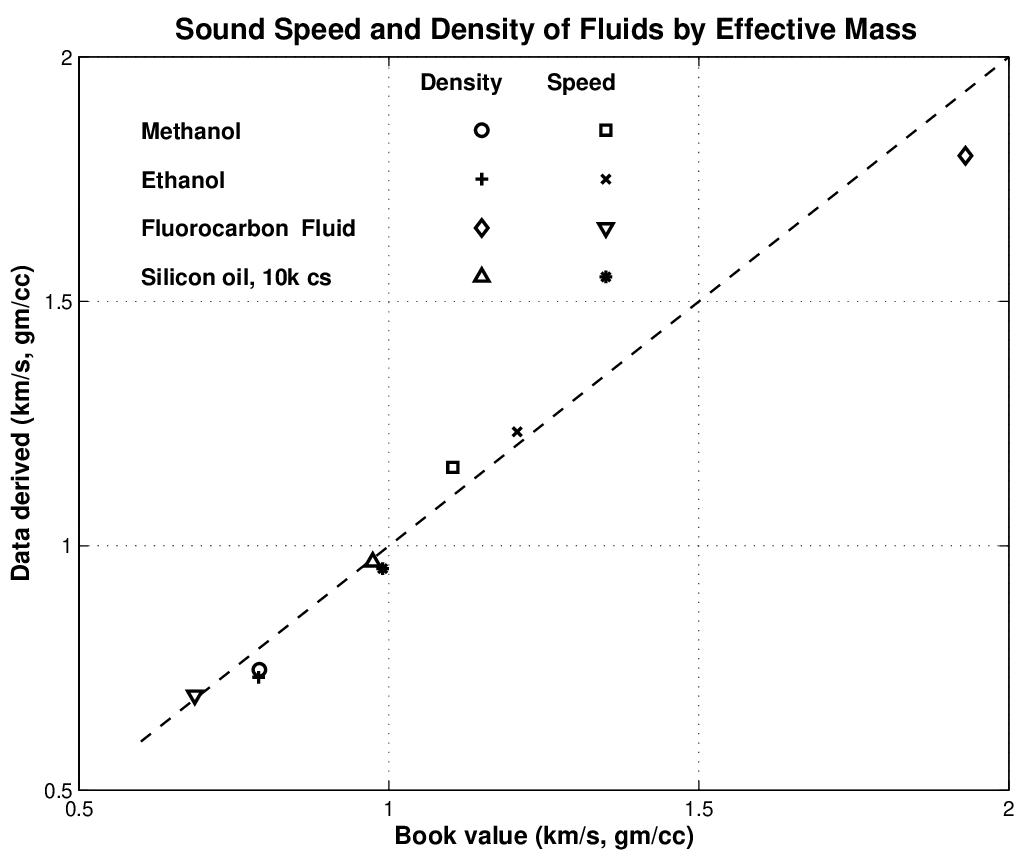}} } \caption {Densities and
sound speeds of various liquids measured with our effective mass
technique and crossplot against those values determined by more
conventional means \cite{selfridge,crc}. } \label{fluids}
\end{figure}

A wide variety of real liquids satisfies the assumptions of Model I.
We have measured $\tilde{M}(\omega)$ for four simple liquids.  In
Figure \ref{fluorinert} we show the results for a common,
commercially available, fluorocarbon, whose chemical formula is
$N(CF_3(CF_2)_4)_3$.  By fitting Eq. \rf{fluid1} to this data, we
are able to extract the density, $\rho$, and the speed of sound: $V
= \sqrt{K_0/\rho}$. These values, measured with our effective-mass
technique for all four liquids, are cross-plotted against those
determined by more conventional means \cite{selfridge,crc} in Figure
\ref{fluids}. As there is a good agreement, both for density and for
sound speed, we conclude that our technique for measuring the
dynamic effective mass $\tilde M(\omega)$ is an accurate one.

\subsection{Granular Media}\label{sec:granmed}

Encouraged by these results, we naively interpret the main resonance
in Figures \ref{F_a_m} and \ref{repro} as being a 1/4 wavelength
resonance of the compressional sound speed i.e. analogous to the
resonances predicted by Model I.  We investigated how this
resonance, $f_g$, shifts to higher frequencies as the filling depth,
$L$, is reduced. Throughout the volume of grains in the cup, the
sound speed must be depth dependent; the gravity-attributed
stiffness is small at the surface and maximum at the bottom.
Nevertheless, we may estimate an effective sound speed in the
vertical direction based on this peak frequency and on the filling
depth of the tungsten particles: \beq{res} v(L) = 4 L f_g(L).
\end{equation}

\begin{figure}[!htbp]
        \centering
        \hbox{ \resizebox{6.0cm}{!} {
        \includegraphics{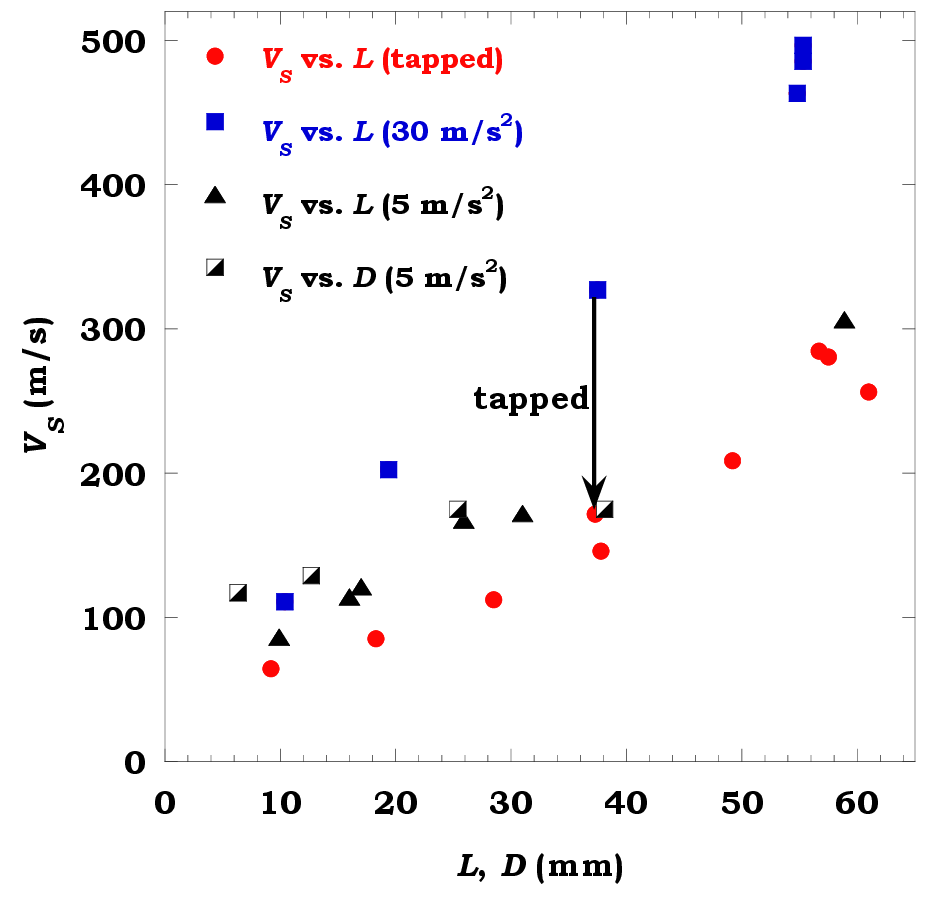}}}
        \caption{(Color online) Effective
speed of sound in tungsten granules as a function of filling depth,
L, for constant cavity diameter, D = 2.54 cm , and as a function of
cavity diameter, filled to a constant depth, L = 3.05 cm.  The
numbers in parentheses refer to the maximum amplitude of vibration
during the packing protocol.} \label{depthdep}
    \end{figure}

Figure \ref{depthdep} shows these estimated speeds as a function of
filling level, L, for a cavity with diameter D = 2.54 cm and for
cavities of differing diameters, all filled to the same level, L =
3.05 cm. For this figure we have filled the cavities with two
different procedures: (1) We vibrate the cavity vertically at 1 (kHz)
with different acceleration amplitudes, as indicated. (2) We fill
simply by ``tapping gently" on the side of the cup.  We note that
the position of the main resonance in the cup can be very different
depending upon the filling technique.  Nonetheless, these results
show the trend of greater speed with greater depth, as expected. The
values we are reporting $\sim$ 100-300 $(m\cdot s^{-2})$ are of the same order of
magnitude as those of other granular media reported in the
literature using other techniques
\cite{cremer,Bourinet2,Varanasi_b,liu,shield, schmidt}.

In one case shown in Figure \ref{depthdep} we first filled the cup
by the vibration measurement and then lightly tapped the side of the
cup.  This caused the grains to pack less tightly and move the main
resonance to a much lower frequency, and thus a much lower apparent
sound speed.

Also in Figure \ref{depthdep} we show our results for the effective
sound speed in cavities of differing diameters, D, filled to a
common depth.  These data provide evidence of a kind of dynamic
Janssen effect in the sense that the side walls support some
fraction of the differential force as an effect of the oscillation.
In order not to confuse the issue with the dynamic Janssen effect
reported previously for that observed in cavities whose walls move
at a constant speed \cite{bertho}, we will refer to the effect in
the present paper as the acoustic Janssen effect.  If the side walls
did not support the effective mass one would expect the effective
sound speed to be independent of cavity diameter.  We return to this
point in our discussions of numerical simulations, below.  Suffice
it to say that the results of Figure \ref{depthdep} rule out the
strict applicability of an interpretation of the main resonance peak
based on Eq. \rf{fluid1}, Model I:  The sound speed is a function of
depth and the material has a shear rigidity. This observation, a
sound speed in the range 100-300 $(m\cdot s^{-2})$, implies the elastic moduli are
in the range $K \sim 1-9 \times 10^7$ (Pa), which is orders of
magnitude smaller than the elastic constants of steel.  Therefore,
the presence of the granular media in the cavity of the resonance
bar contributes negligibly to the stiffening of that bar, as we have
assumed.

\begin{figure}
\hbox{ \resizebox{8cm}{!} { \includegraphics{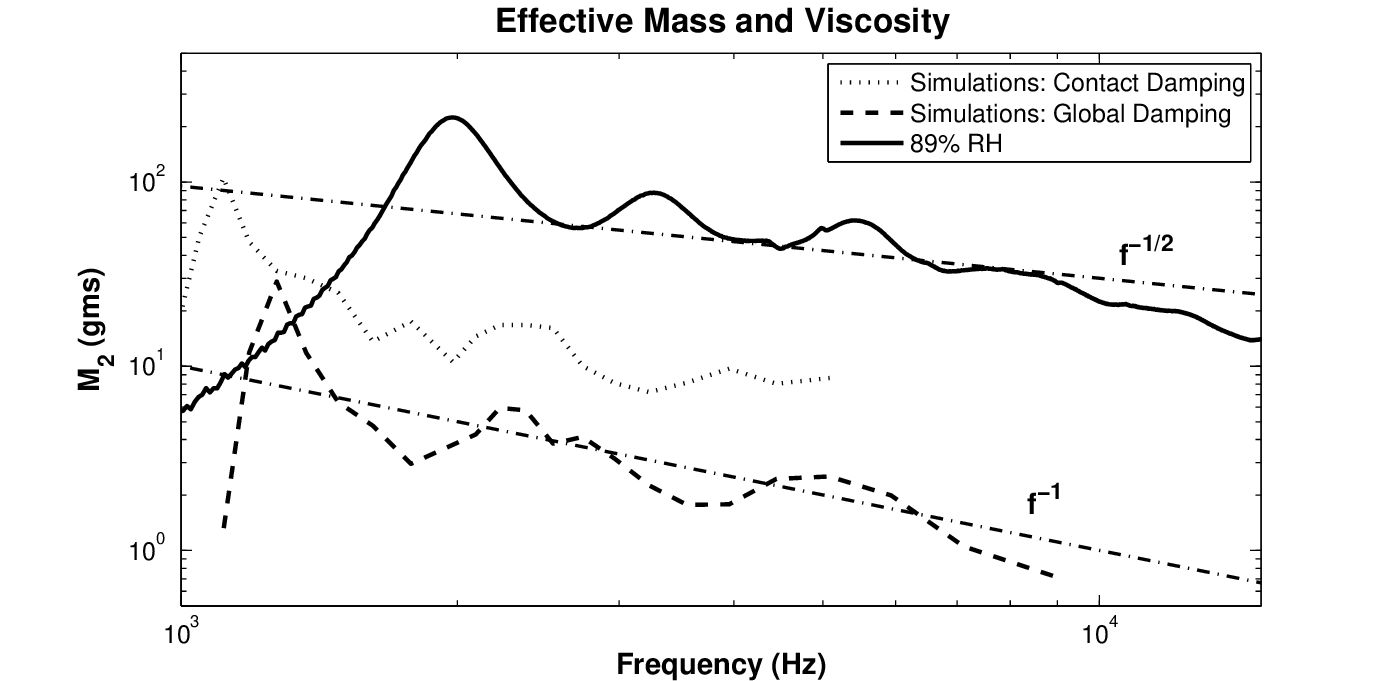}} }
\caption {$M_2(\omega)$ vs. frequency indicating the possible effect
of a macroscopic viscosity in the experimental data.  The simulation
based on contact dampening is suggestive of a macroscopic viscosity
$M_2 \propto f^{-1/2}$ whereas that based on global dampening $M_2
\propto f^{-1}$ is not.  The simulation results have been shifted
vertically.} \label{viscosity}
\end{figure}

Because at least some of the effective mass is borne by the side
walls of the cavity, we may estimate the effective viscosity of the
granular medium based on the high-frequency tail of the data, such
as in Figures \ref{F_a_m} and \ref{repro}.  According to Model II,
the high frequency tail of $\tilde{M}(\omega)$ should be
proportional to $\omega^{-1/2}$, as in Eq. \rf{hfluid}, when the
viscous skin depth is small: $\delta(\omega) \ll a$. Roughly
speaking, this is seen in the data plotted in Figure
\ref{viscosity}.  (The data fit better to $f^{-1/2}$ than to
$f^{-1}$ although not overwhelmingly so.) Taking into account the
prefactors, we conclude that the granular medium has an effective
acoustic viscosity $\eta_{\rm eff} = [3 - 10] \times 10^4$ Poise.
This value of $\eta_{eff}$ implies a value of the viscous skin depth
$\delta(10 kHz) \approx 3 mm$ which is both small compared to the
cup radius and large compared to the grain diameters thus satisfying
two of the underlying assumptions in Eq. \rf{hfluid}.  This value is
just an order of magnitude estimate as the fluctuations around the
$f^{-1/2}$ behavior are quite large and Model II is, of course, an
oversimplification and an overestimation of the effects of the sides
of the cavities.  It is clear that, while the data exhibit features
of both Model I and II, neither model captures the whole story.
 Nonetheless, if our estimate of the macroscopic viscosity is
approximately correct that would imply that the crossover from $M_2
\propto \omega^{-1/2}$ to $M_2 \propto \omega^{-1}$ [i. e. Eq.
\rf{hflim2}] occurs in the frequency range $[2-8] \times 10^6$ (Hz).
This crossover happens when the macroscopic viscous skin depth
approximately equals the inter-particle separation, $\delta = b$, as
discussed in connection with Figure \ref{one_D_contact}.  If this
crossover happens at all, it would seem to be well outside our
experimentally accessible measurement range.

In a similar vein, we may conceptualize the effective elastic
constants as $\tilde{K}(\omega) = K_o -i\omega \eta$ from which we
deduce that a crossover from  elastic-dominated to viscous-dominated
behavior occurs at a frequency $\omega_{e-v} \sim K_o/\eta$ which is
approximately 15 (kHz) for the granular systems we have studied.  We
may very well be seeing this crossover at the higher end of our
frequency range in Figure \ref{viscosity}.

To summarize this subsection we have found some evidence that there
is a an effective macroscopic viscosity in the sense of Eq.
\rf{hfluid} being approximately correct.  This, in turn, suggests
that we should not see behavior implied by Eq. \rf{hflim2} unless
the frequency is much higher than we are investigating; we don't see
that behavior, much less that suggested by Eq. \rf{hflim3}.  Even at
our highest frequencies the granular system is undergoing collective
oscillation, albeit a complicated one: It is {\it not} the case,
even at our highest frequency, that only those grains in contact
with the walls are contributing to $\tilde{M}(\omega)$.

\section{Simulations}\label{sec:simulations}

The toy models are illuminating as far as they go but to obtain a
deeper understanding of the dampening mechanism on a microscopic
level we have performed molecular dynamic simulations of $\tilde
M(\omega)$.  Here, we consider the much simpler system of spherical
beads confined in a rectangular box.

In our simulations a static packing at a pre-determined pressure is
first achieved by methods previously described \cite{mgjs}, and then
we incorporate walls, friction, and the force of gravity. The
simulations consider the typical Hertz-Mindlin contact forces for
the normal and tangential components, respectively, and the presence
of Coulomb friction between the particles characterized by a
friction coefficient $\mu$.  See Ref. \cite{mgjs} and references
therein for a discussion of the underlying physics of these contact
forces.

Microscopic dampening is provided by two principal mechanisms of
dissipation: (a) Local dampening, in which the force is proportional
to the relative velocity between two contacting grains
\cite{poschel2,wolf-md}.  Examples of this form of dampening include
intrinsic attenuation due to asperity deformation \cite{poschel2},
and wetting dynamics within the liquid bridges between adjacent
grains as they move relative to each other \cite{crassous1},
\cite{crassous2}.  (b) Global dampening, in which there is a
presumed rotational and translational drag due to a viscous fluid,
such as air, which is assumed to move with the walls of the cavity.
See \cite{rayleigh, thornton, wolf-md}.   Within the context of Eqs.
\rf{EoM2} and \rf{EoM3} the global dampening approximation is that
the particles not in direct contact with the walls still experience
a drag effect such that ${\bf K}_{iw} = -i\omega {\bf B_w}$, where
${\bf B_w}$ is the same for each grain. The conclusions we draw from
our numerical simulations are relatively insensitive to the assumed
values for the dampening parameters, either global or local.  We
plot some typical results in Figure \ref{viscosity} where we indeed
see that the local dampening hypothesis implies $M_2 \propto
\omega^{-1/2}$ and the global dampening is more consistent with $M_2
\propto \omega^{-1}$. The data are quite noisy and the distinction
between $f^{-1/2}$ and $f^{-1}$ behavior is not clear-cut, although
the fit to $f^{-1/2}$ is somewhat better.  In Section
\ref{sec:granmdehum} we demonstrate quite unequivocally that contact
damping is the operative mechanism in the experimental data.

Results are displayed in Figure \ref{simulations} for a system in
which there is assumed to be only local contact dampening. The two
cases shown correspond to  normal and transverse contact forces
(top) and normal forces only (bottom).  The fundamental resonance
and the broad high-frequency tail are clearly evident.  We show,
separately, the contribution to $\tilde{M}(\omega)$ from the bottom
as well as from each of the side walls.  The conclusions we draw
from the numerical modeling are:

\begin{figure}
\centerline{\hbox{ \resizebox{6cm}{!}{\includegraphics{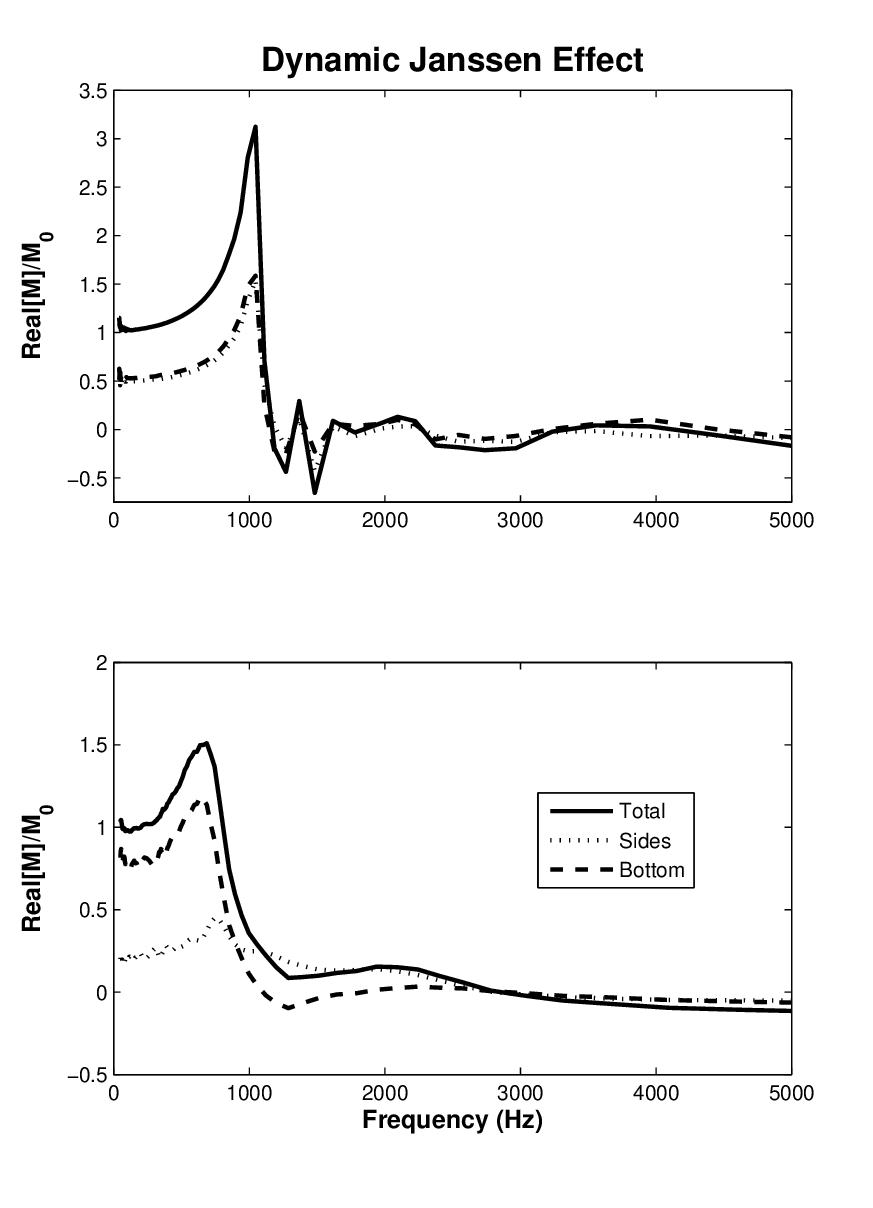}}
}}
 \caption
{Acoustic Janssen effect as seen in molecular dynamic simulations.
The real part of the computed effective mass, normalized to the
static mass, is plotted versus frequency: (Top) $\mu=0.5$, (Bottom)
$\mu=0$. Although both simulations clearly show the resonance, in
the former approximately one-half of the effective mass is supported
by the side walls and one-half by the bottom. } \label{simulations}
\end{figure}

1. An acoustic Janssen effect reveals that Models I and II (Section
\ref{sec:contmod})  are equally important to an understanding of the
dynamics. Figure \ref{simulations}a shows the results of the
effective mass at the bottom and on the walls of the cavity, as a
function of the frequency, for a system with friction coefficient
$\mu=0.5$.  For all frequencies, we find that approximately one-half
of the mass is held by the bottom and the other half by the side
walls.  In this sense, one cannot make the distinction between the
simplified models.  Our depiction of an effective sound speed as a
function of filling level, Figure \ref{depthdep}, is really just a
manner of speaking.  When the friction is switched off ($\mu=0$,
Fig. \ref{simulations}b), almost all the weight is supported by the
bottom walls of the cavity, as expected, since the effective shear
modulus becomes negligibly small \cite{mgjs}; therefore the Model II
effect essentially disappears. (A small component of the weight is
held by the walls because they are made of glued particles in the
simulations.) Not surprisingly, we still see the resonance peaks as
predicted by Model I.

We note that our results for the effective sound speed measured in
cavities of different diameters but filled to the same depth,
plotted in Figure \ref{depthdep}, provide an experimental indication
of an acoustic Janssen effect.  If the side walls did not support
the effective mass one would expect the effective sound speed to be
independent of cavity diameter.

2. Simulations allow us to differentiate between possible different
microscopic origins of dissipation.  We find that either global or
local dampening can capture the main features of the experiments:
the main resonance peak, as in Model I, and a broad background, as
in Model II.  However, the high-frequency asymptotic behavior for
large $\omega$ is very different for the two mechanisms.  Global
dampening predicts $\tilde{M}(\omega) \sim i \omega^{-1}$, as per
Eq. \rf{hflim2} where $B_{iw}$ is nonzero for every particle.
Roughly speaking this can be seen in the result plotted in Fig.
\ref{viscosity}. On the other hand, as long as the viscous skin
depth $\sqrt{2 \eta_{\rm eff}/\rho \omega}$ is large compared to the
grain size but small compared to the cup radius, then contact
dampening predicts $\tilde M(\omega) \sim (i/ \omega)^{1/2}$, as in
Model II.  This trend is seen in the numerical simulations based on
contact dampening plotted in Fig. \ref{viscosity}.  Although the
simulation result is noisy, we may conclude that there is an
effective viscosity, as is seen in the experimental results. Contact
dampening can be caused by viscoelasticity of the grains themselves
or it can be induced by liquid bridges at the contact points
\cite{ocon,damour, crassous1, crassous2}. We are inclined to suppose
it is the latter that dominates in our samples and this has
motivated us to consider the effects of humidity on our results,
both for $\tilde{M}(\omega)$ and for the resonances in the bar.  We
present our results in the next section. For reasons which will
become apparent, we need to develop a quantitative theory of the bar
resonances that goes beyond the perturbation theory, which we also
describe in the next section

\section{Detailed Theory of Flex Bar Resonances}\label{sec:timo}
The perturbation theory described in Section \ref{sec:perth} is
informative but in some situations  $\tilde{M}(\omega)$ takes on
very large values - larger even than the mass of the bar itself,
thus invalidating the assumption that $\tilde{M}(\omega)$ is a small
perturbation. Moreover, there are resonances seen in the loaded bar
which are primarily resonances within the granules themselves;
perturbation theory is not able to make any prediction about these
resonances. For situations such as these it is necessary to go
beyond the perturbation theory and use a more complete theory.  For
the case of a rectangular bar the flex mode resonances, including
the effect of the granular medium, can be computed directly.  This
technique is described in Appendices A and B.  Basically, the theory
treats the bar as a one dimensional object for which flex waves are
described by the so-called Timoshenko beam theory \cite{timoshenko}.
The extra compliance near the center of the bar, due to the
existence of the cavity, is modeled by a single parameter, $\xi$,
whose value is set by the resonance frequency when the cavity is
empty. There are no other adjustable parameters. As before, though,
we treat the effective mass of the grains $\tilde{M}(\omega)$ as a
localized point density.

At each (complex-valued) frequency there are two left-going and two
right-going waves.  The amplitudes of these four components are
determined by the requirement that four homogeneous boundary
conditions must be satisfied.  The normal-mode condition is that the
determinant of these coefficients must vanish; we search,
numerically, for the complex-valued normal-mode frequencies,
$\{\omega_R\}$, at which the determinant vanishes.

We compare the results of this theory, first against resonance bar
data taken when the cavity is filled to different depths with
various liquids, and then to data taken when the cavity is filled
with granular media, under conditions of differing humidity.
\subsection{Liquids}
\begin{figure}
\hbox{ \resizebox{8cm}{!} { \includegraphics{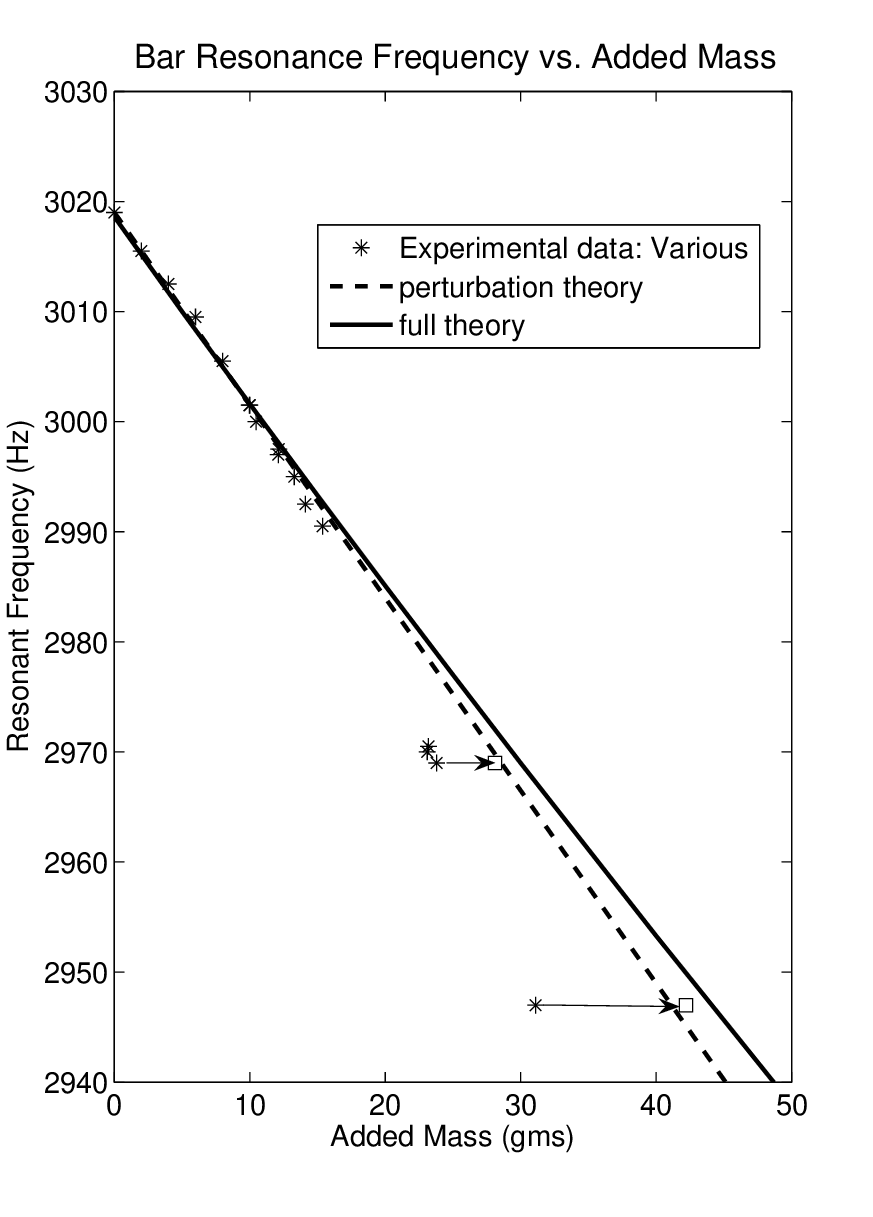}} }
\caption {Resonance frequency of steel bar as a function of added
mass in the cavity, for various liquids, filled to various depths.
The solid curve is the prediction based on the theory described in
Appendix A.  The dashed  curve is the perturbation theory thereof
Eq.\rf{pertheory}.  The two data points indicated with squares
represent the effective mass of those liquids evaluated at the
resonance frequency, using Eq. \rf{fluid1}.} \label{baraddedmass}
\end{figure}

In Section \ref{sec:liquidEM} we showed how the effective mass
measurements on simple liquids gave results in general agreement
with the predictions of Model I, Eq. \rf{fluid1}.  It is natural to
inquire whether the theory described in Appendix A can accurately
predict the frequency shift of a resonant bar whose cavity is filled
to varying depths with such a simple liquid.  In Figure
\ref{baraddedmass} we show the measured results of the flexural
resonance frequency of the bar when the cavity is partially filled
with the liquids considered in connection with Figure \ref{fluids}.
The solid curve represents the predictions of the theory described
in Appendix A, assuming the effective mass within the cavity is
frequency independent.  Over the range of added mass values, the
full theory is nearly the same as the predictions of the
perturbation theory, Eq. \rf{pertheory}. We have determined $I_1
\approx  2.1$ for the fundamental mode in our bar, reflecting the
fact that the displacement at the center is much larger than the
average RMS displacement.  For large enough values of added mass the
resonance frequency predicted by the full theory asymptotes to a
finite frequency, reflecting the fact that even if the center of the
bar is pinned, the two arms may still oscillate freely.  The
difference between the full theory and the perturbation theory is
quite large for added masses on the order of 500 grams, which is off
the scale of Figure \ref{fluids}. The data mostly lie on the
theoretical curve except for the data points greater than 20 g,
which correspond to a fluorocarbon fluid.  For these data points it
is simple enough to estimate $\tilde{M}(\omega)$, evaluated at the
measured resonance frequency, using Eq. \rf{fluid1}.  With this
correction to the added mass all the data for the simple fluids now
lie nearly on the theoretical curve, which gives us confidence in
our approach.

\subsection{Granular Media - Humidity Effect}\label{sec:granmdehum}
In order to elucidate the physical origin of the dampening mechanism
we have undertaken a controlled study of the effects of humidity on
these systems.  Both the shaker cup and the resonant bar are filled
with the same amount of grains, by weight, of tungsten particles.
They are packed into their respective cavities using the mechanical
compaction protocol described in Section \ref{sec:samphand}. Both
the shaker apparatus and the resonant bar are placed in a
hermetically sealed glove box.  The humidity is controlled by
placing an open pan of salt-saturated water inside, as well.  We
have used different salts in the water as a means of controlling the
humidity.  We use a low power fan to provide a continuous flow of
air throughout the container. The temperature is held at T =
$26.5^o$ C.   The glove box sits on a vibration isolation table. The
motivation here is to allow for equilibration of humidity on a
reasonable time scale and to not allow for extraneous vibrations to
dislodge the grains in the cavities.

\begin{figure}
\hbox{ \resizebox{7cm}{!} { \includegraphics{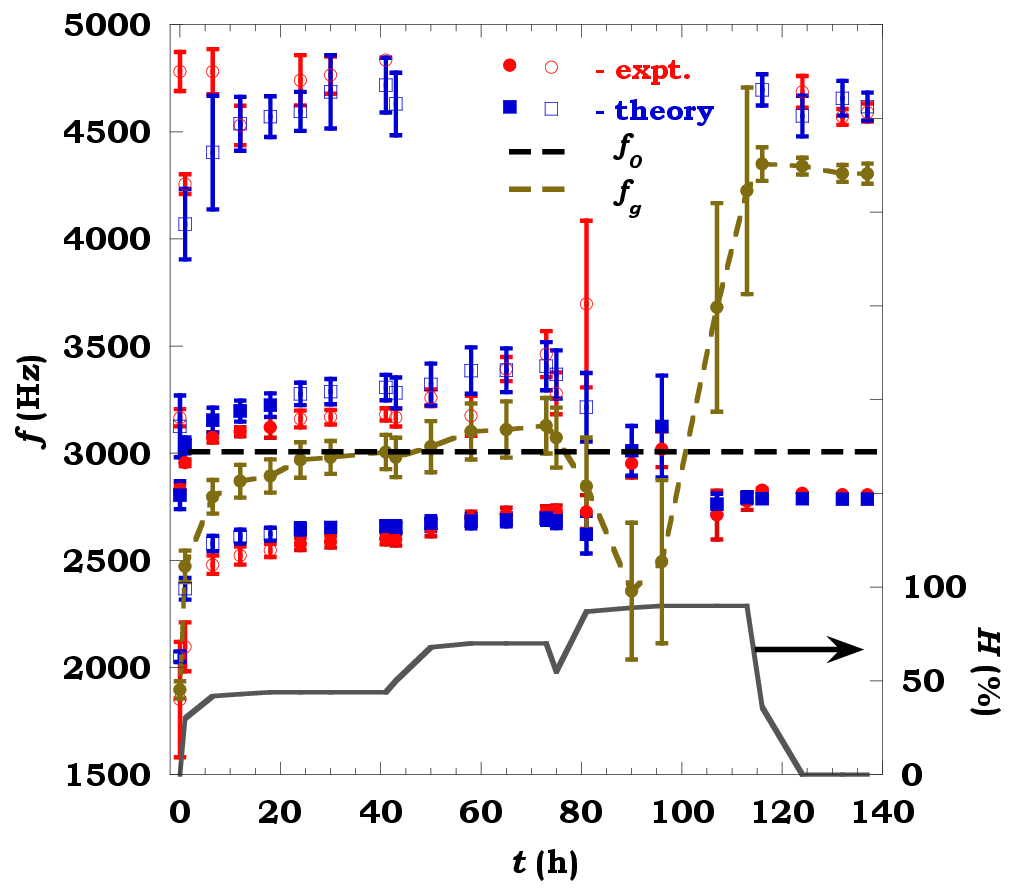}} }
\caption {(Color) Comparison of measured and predicted resonant bar
characteristics at various humidities.  The ``error bars" indicate
the width of the resonances, as described in the text.  The filled
symbols refer to the dominant mode in the system, which is
predominantly ``bar-like". The horizontal dashed line, $f_0$, is the
resonance frequency in the unloaded bar. Its Q is so high that the
width does not show on the scale of the figure. $f_g$ shows the
position and width of the main resonance seen in the effective mass
data, some of which is show in Figure \ref{Mvstime}. The solid curve
is the relative humidity in the glove box vs. time.}
\label{glove_box}
\end{figure}

In Figure \ref{glove_box} we show a comparison of the measured and
the computed resonance frequencies in the bar as a function of
elapsed time.  We also show how the humidity is changing, as we swap
one salt-saturated solution for another. (The ``zero" humidity cases
are accomplished with a desiccant.)  There are several different
resonance frequencies that were measurable.  For each of them we
have determined the complex-valued resonance frequency,
$\tilde{f_R}$, using the procedure described at the end of Section
\ref{sec:resbar}.  The width of each resonance is indicated with the
error bars. i.e. What is plotted is ${\rm Re}(\tilde{f_R})$ and
${\rm Re}(\tilde{f_R})\pm {\rm Im}(\tilde{f_R})$.  Using the
simultaneously measured $\tilde{M}(\omega)$ in the shaker cup, we
are able to compute the expected (complex-valued) resonance
frequencies in the bar, using the theory described in Appendices A
and B.  These computed resonance frequencies are depicted in the
same manner as the measured ones, in Figure \ref{glove_box}.  The
horizontal dashed line is the resonance frequency in the unloaded
bar, $f_0$.  Its Q is so high ($\sim 1000$) that the width (a few
Hertz) does not show on the scale of Figure \ref{glove_box}; see Figure
\ref{resonant_bar}, top. The other dashed curve in Figure
\ref{glove_box} plots the position and width of the main resonance
frequency seen in $\tilde{M}(\omega)$, $f_g$.

\begin{figure}
\hbox{ \resizebox{7cm}{!} { \includegraphics{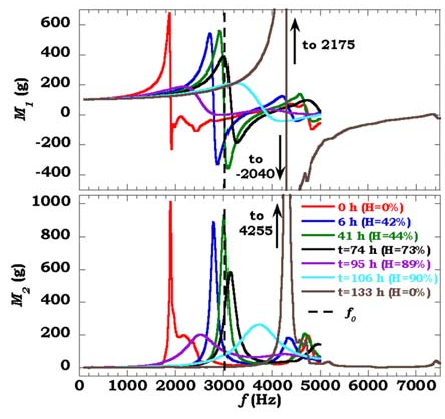}} }
\caption {(Color) Evolution of dynamic effective mass as the
humidity is changed, as per the humidity data in Figure
\ref{glove_box}.} \label{Mvstime}
\end{figure}

A few selected plots of $\tilde{M}(\omega)$ are shown in Figure
\ref{Mvstime}.  The main resonance in $\tilde{M}(\omega)$ is clearly
visible in Figure \ref{Mvstime} but, in fact, there is another
resonance of smaller amplitude, around 4500 (Hz) for the 43\% and 73\%
humidity cases, which is barely visible in Figure \ref{Mvstime}.

There are several interesting features to the data of these two
Figures.  First, from either Figure \ref{glove_box} or Figure
\ref{Mvstime} we see that the position of the main resonance within
$\tilde{M}(\omega)$ starts at a low [$\sim 2000$ (Hz)] frequency when
the humidity is initially zero.  It then increases as the humidity
increases to 43\% then again to 73\%.  Concomitantly, the width of
that resonance also increases; the system of loose particles is
becoming more dampened with increasing humidity.  Second, there is
generally quite good agreement between theory and experiment as to
the position and widths of the several different resonances in
Figure \ref{glove_box}. Third, although the strongest resonance in
the bar around 3 (kHz) dominates, there are also one or two other
resonances, which we associate with the dynamics within the granular
medium.

We have noticed that, of the two resonances seen in the bar in the
region 2000-3500 (Hz), one of them has a significantly larger
amplitude than the other, as is clear from the example shown in
Figure \ref{resonant_bar}, which corresponds to t = 41 hours in
Figures \ref{glove_box} and \ref{Mvstime}. This ``stronger"
resonance is labeled with filled symbols, for both the measured and
computed resonances, in the legend of Figure \ref{glove_box}; the
others are labeled with open symbols. Our interpretation is that the
two ``bare" modes, the resonance in the empty bar at $f_0$, and the
resonance in $\tilde{M}(\omega)$ at $f_g$, have coupled to each
other to yield two hybrid modes in the bar + grains system, an upper
branch and a lower branch.  The resonance with the larger amplitude
corresponds to a mode which is predominantly bar-like.  When $f_g <
f_0$, at early times, the real part of the effective mass is
negative in the vicinity of $f_0$. Although the theoretical
calculations were done with the complete theory of Appendix A, it is
useful to consider the qualitative predictions of the perturbation
theory.  According to the predictions of Eq. \rf{pertheory}, one may
expect the resonant frequency in the bar should {\it increase}
relative to that of the empty bar due to this negative mass loading,
and this is just what we see in the upper branch.  At slightly later
times, when the humidity is increased to $\sim 70\%$, $f_g$ has
increased above $f_0$, with the result that the real part of
$\tilde{M}(\omega)$ is now positive, in the region around $f_0$.
Now, according to the perturbation theory, the resonance frequency
in the bar should {\it decrease} and this, too, is just what we see.
An analogous behavior was seen in the data of Kang {\it et al}
\cite{Kang} who considered the behavior of the resonance frequency
of a clamped plate as more and more loose beads were placed atop it.
Our view of their experiments is that for low values of
mass-loading, the resonance frequency, $f_g$, in their grains is
higher than the resonance frequency of their plate. Thus, the real
part of their effective mass is positive and the resonance frequency
is initially lowered, relative to the unloaded value.  If the depth
of loading is large enough, $f_g$ drops below the plate resonance
frequency and the real part of $\tilde{M}(\omega)$ can become
negative.  This is our explanation of why the observed resonance
frequency may then increase, relative to the unloaded value.

Although the frequency of the upper branch in Figure \ref{glove_box}
is relatively slowly varying, it very abruptly changes character,
from ``bar-like", at the early times, to ``grain-like" later.  The
lower branch shows the opposite behavior. This aspect of the
response of the system will be analyzed in detail elsewhere
\cite{Valenza}.

When the system is at 90\% humidity, it behaves quite differently.
(There is an initial drop in humidity which is an artifact of our
procedure for swapping in and out the salt-saturated solutions.) The
main resonance in $\tilde{M}(\omega)$ initially {\it drops} to
around 2200 (Hz) but, more significantly, it becomes very much
broadened relative to that at the lower humidities, cf. Figure
\ref{Mvstime}.  We do not understand the initial drop in the
resonance frequency; conceivably it is related to the initial drop
in humidity.  At this humidity, we were able to locate only one
resonance in the experimental bar data and we were also able to
locate only one in the theoretical computations. As time goes on,
maintaining the humidity in the glove box constant, we see from
Figures \ref{glove_box} and \ref{Mvstime} that the main resonance in
$\tilde{M}(\omega)$ increases with time. It increases with time for
the lower humidities, too, though it is most pronounced for 90\%
humidity.  Previously \cite{Hsu}, we analyzed this aspect of the
data in the context of an assumed distribution of asperity heights
within the contact region.  With time more and more of the
energetically favorable asperity regions become filled with
condensate, assuming they are smaller than the Kelvin radius,
Eq.\rf{Kelvin}, below. Under a fairly broad set of assumptions, this
lead to the prediction that the contact stiffnesses would increase
approximately logarithmic in time, which is what was observed.  The
data sets analyzed in Reference \cite{Hsu} were, however, prepared
using the vibration protocol described earlier in Section
\ref{sec:samphand}, unlike the data in Figures \ref{glove_box} and
\ref{Mvstime}.  Nonetheless, we believe that this is what is
happening:  At any given level of humidity, there is a relatively
slow process of equilibration leading to a relatively slow increase
in the resonance frequency within $\tilde{M}(\omega)$.

Finally, when the salt solution is replaced with a desiccant,
causing the humidity to drop to essentially zero, there is a very
dramatic change in the properties of the granular medium.  According
to Figure \ref{Mvstime} the main resonance has shifted up to a very
high frequency, $\sim 4300$ (Hz). It has become extremely sharp, and
extremely strong.  Concurrently, in the bar there are two distinct
resonances, the one primarily within the grains, around 4500 (Hz), and
the other primarily in the bar, around 2800 (Hz).  We note that both
of these are significantly sharper (higher Q) resonances than their
counterparts in the humid stages.  We note that theory and
experiment are in general agreement as to the level of attenuation
of the various modes.  Specifically, the attenuation of the bar
resonance [$\sim 3000$ (Hz)] is much reduced in the dried out state
relative to that in the humid states.

\begin{figure}
\hbox{ \resizebox{7cm}{!} { \includegraphics{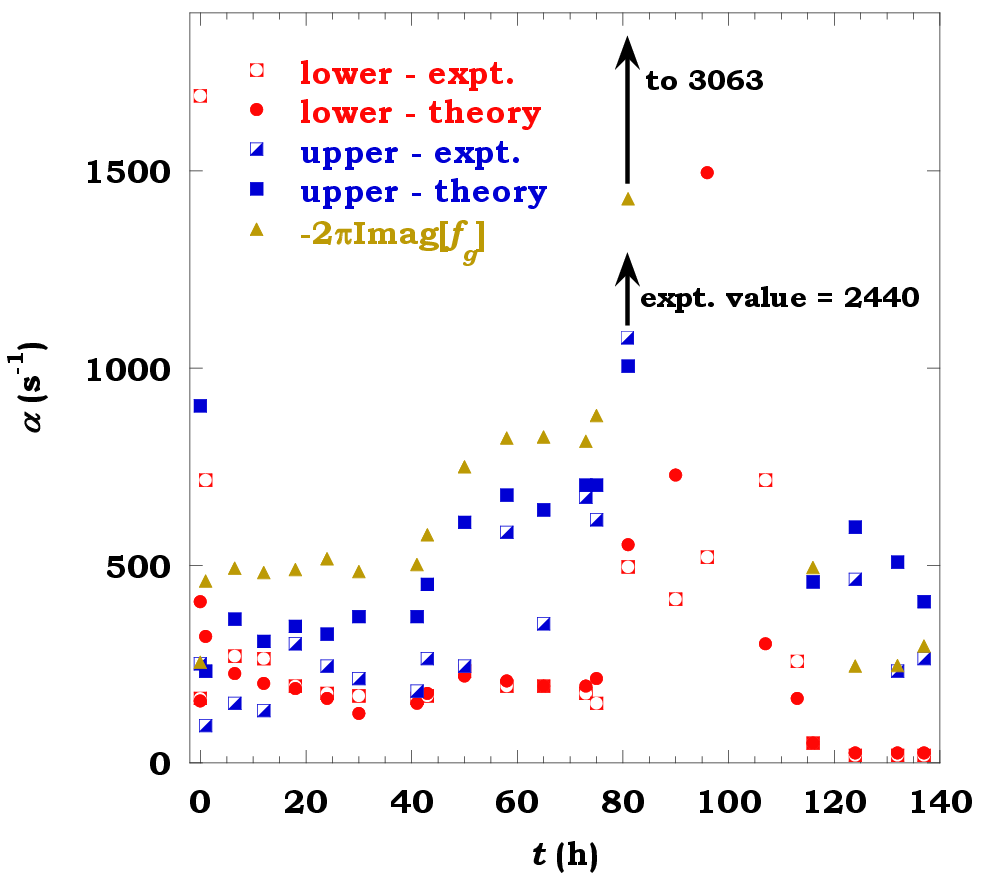}} }
\caption {(Color) Decay rate, $\alpha \equiv -{\rm Im} \omega_R$, of
the various normal modes depicted in Figure \ref{glove_box}. Here,
``upper" and ``lower" refer to the two branches with resonance
frequency near 3 (kHz).} \label{decay_rate}
\end{figure}

We don't really know why we see two resonances in the dry cases and
one, two, or three in the humid cases, depending, but this seems to
be the situation both for the theory and the experiment.  In the
theory, for 90\% humidity, one may speculate that there is an
additional resonance or resonances, presumably related to the
apparent resonance seen in Figure \ref{Mvstime}, but that it/they
appear for very large values of the dampening rate i.e. well off the
real axis. However, if they exist we have been unable to locate such
resonances with our theoretical machinery no matter what value we
use for seeding the mode search, as described in Appendix A.

In order to analyze the effectiveness of the granular medium in
attenuating the modes in the bar, at the various humidities, we
re-plot part of the data of Figure \ref{glove_box} in Figure
\ref{decay_rate}.  Here, $\alpha \equiv \;{\rm Im} [\omega_R]$
represents the decay rate of the amplitude, $A$, of each mode, in
the time domain, viz: $A \propto e^{-\alpha t}$.  We are quite
pleased with the general level of agreement between theory and
experiment, the exceptions occurring when the system is still
equilibrating to humidity changes, from 80 to 110 hours.  We note
also that the agreement is generally better for the ``bar-like" mode
than it is for the ``grain-like" mode.  The decay rate of the
unloaded bar is so small, $\alpha$(unloaded) = 9.4 $(neper\cdot s^{-1})$, that it
appears to be zero on the scale of Figure \ref{decay_rate}. We see
from the Figure that there is a monotonic effect of humidity on the
decay rate of the main resonance within the granular medium, as
shown in gold. Such is {\it not} the case for the dampening
coefficient of the mode which is predominantly bar-like in the
combined system. Even though the humidity varies from 43\% to 70\%
in the first 80 hours, the measured dampening rate of this mode
exhibits only a small, non-systematic, variation.  This variation
is, however, captured in the theory. At $ t = 80$ hours, when the
humidity is raised to 90\%, the measured and the predicted dampening
rates of this mode both increase enormously. When the salt-saturated
solution is replaced with a desiccant the measured and computed
dampening rates drop to very small values, much smaller than when
the system was originally in the dry state. (We note that the grains
in the shaker cup seem to equilibrate with humidity changes more
rapidly than do the grains in the bar, but eventually they tend to
the same state. Perhaps this is a consequence of the different flow
patterns of air in the vicinity of the two cavities in the glove
box.)

\begin{figure}[!htbp]
        \centering
        \includegraphics[width=0.85\columnwidth]{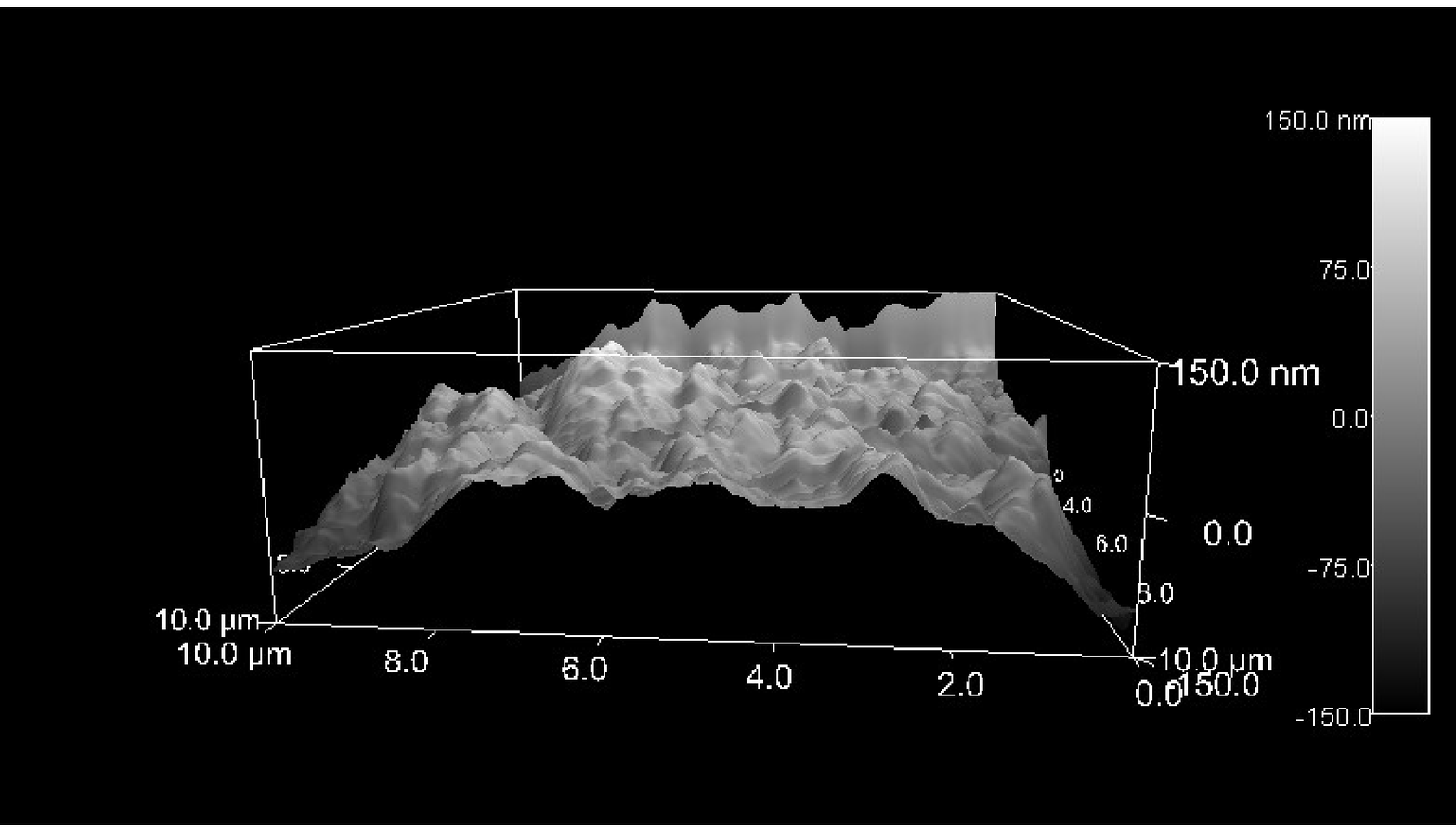}
\caption{AFM image of the surface texture of a tungsten granule
showing asperities on the 10-20 nm scale.  Note the differing
vertical and horizontal scales.} \label{surface_roughness}
    \end{figure}

Our observed humidity effect is very much analogous to that seen in
the results of D'Amour {\it et al.} \cite{damour} who deduced very
similar behavior from their measurements of the changes in resonance
behavior of a quartz plate whose top surface was covered with a
monolayer of beads.  They concluded that the {\it in situ} drying of
the contacts between the beads and the quartz causes them to stick
over a larger microscopic surface area, causing the asperities on
the surfaces in the contact zone to be partially crushed. This, in
turn, causes the stiffness of the contact ``springs" to increase and
the dampening to decrease.  A rough analogy would be when one soaks
tissue paper with water it is easy to make it stick to a vertical
surface; the mess is held up by the force of surface tension.  If it
is allowed to dry, the force of adhesion to the wall actually
increases, even though the water is gone.  That is because as it
dries it also deforms, allowing a large area of solid-on-solid
contact to develop. This, basically, is what we see in our own
measurements.  An AFM image of the surface of one of the tungsten
granules in Figure \ref{surface_roughness} shows these asperities on
the scale of 10-20 nm.  Capillary condensation within these
asperities occurs when the Kelvin radius, $r_m$, is approximately
equal to the asperity size where \beq{Kelvin} r_m = -\frac{2 \gamma
V_L}{RT\ln(H)} \:= -\frac{1.04 \:{\rm nm}}{\ln(H)}\:\:\:,\eeq where
$\gamma$ is the liquid vapor surface tension, $R$ is the gas
constant, $T$ is the absolute temperature, $V_L$ is the molar
volume, and $H$ is the relative humidity \cite{GreggSing}. The
numerical value in the second equation is that which is appropriate
to water at $26.5^o$ C. Thus, at a relative humidity of 90\% the
Kelvin radius is approximately 10 nm, which is large enough that a
single liquid bridge engulfs all the asperities in the contact
region between neighboring particles.

We conclude, therefore, that the dominant mechanism of acoustic
dampening in granular media is due to the adsorbed film of water
which exists on the particles, an example of contact dampening. In
particular, the film in the region between two contacting grains
undergoes a shear deformation due to the relative motion of the two
grains and thus it gives rise to dampening due to the viscosity of
the water in the film.  Conceptually this may be thought of as an
example of the ``squirt mechanism" \cite{squirt} which is operative
as the dominant attenuation mechanisms in sedimentary rocks at
ultrasonic frequencies.  In that case there is clear evidence that
the pore fluid is squeezed in and out of microscopic cracks which
exist within the cement material holding the grains together.

The data in Figure \ref{decay_rate}, especially the low attenuation
in the second dry state, clearly rule out global dampening or
intrinsic dampening within the Tungsten as being significant
mechanisms for dampening.

It is also clear from Figures \ref{glove_box} and \ref{Mvstime} that
the adsorbed films of water at the grain-grain contacts is a
significant mechanism for stiffness of the inter-granular spring
constants. However, the sound speed data in Figure \ref{depthdep}
indicate that something like a Hertz-Mindlin contact forces must
also be operative:  Why else would the effective sounds speeds be
dependent upon the cavity dimensions? Because of the odd shape of
our tungsten granules it is difficult to analyze the relative
contributions to the grain-grain stiffness due to each of these
mechanisms.  For the simpler geometry of a random close packing of
spherical glass beads it is possible to do this estimate, precisely
because of the data of D'Amour {\it et al.} \cite{damour} on a
monolayer of beads lying on a vibrating quartz substrate.  From
their published data, their Figure 1 and Equation (5), it is simple
enough to estimate the strength of the spring constants ($\kappa$ in
their notation, not to be confused with our use of $\kappa$ in our
Eq. \ref{fluid2}) between the beads and the quartz plate.  We find,
from their data, \beq{range} \kappa({\rm Single \:Bead: \:Exp.})
=[1-7] \times 10^4\:({\rm N\cdot m^{-1}})\:\:\:, \eeq depending upon the
humidity, in their experiments. These values are much larger than
what one might expect for Hertz contact theory in which the weight
of the bead provides a static compression and subsequent stiffness
of the contact. Equations (1-8) of Norris and Johnson \cite{NJ} in
which the normal force, {\it N}, is the weight of a glass bead,
gives \beq{Hertz1} \kappa ({\rm Single \: Bead: \: Hertz}) = 4.6
\times 10^3 \:({\rm N\cdot m^{-1}})\:\:\:. \eeq In doing this estimate we
assumed the relevant radius of the contact is the radius of the bead
itself, 100 $\mu$m; had we assumed a radius equal to that of a
typical asperity on the surface of the bead, the computed stiffness
would be orders of magnitude smaller than Eq. \rf{Hertz1}. This
estimate is further confirmation of the conclusions of D'Amour {\it
et al.} \cite{damour}, that the stiffness of the contacts is due to
surface forces.  If, however, one considers the contact stiffness of
glass beads at a depth of 2.54 cm, compressed by the weight of the
beads above, one may estimate the expected stiffness predicted by
Hertz contact theory to be \beq{Hertz2} \kappa ({\rm Hertz \: at \:
Depth}) = 2 \times 10^4 \:({\rm N\cdot m^{-1}})\:\:\:. \eeq

Thus, we may say that the contact stiffness for grains near the top
surface of a granular-filled cavity - virtually any granular-filled
cavity - is due to humidity mediated surface forces. However, for
those grains a few centimeters deep Hertz-Mindlin contact theory is
also important.

We conclude this subsection by pointing out that the measurement
technique of D'Amour {\it et al} \cite{damour} is very much
analogous to our measurements on the flex bar.  In both cases one is
looking at changes in the resonance frequency of some system due to
the, relatively small,  perturbation of the granular medium.  In
addition our measurements of $\tilde{M}(\omega)$ represent a more
direct measurement of the underlying physics of the granular medium.
In this regard the humidity effect on $\tilde{M}(\omega)$ is quite
strikingly apparent, over a wide frequency range as shown in Figs.
\ref{glove_box} and \ref{Mvstime}.  Although the dampening factor
for the bar under humid conditions is much greater than that dry,
the differences in the main bar resonance are not nearly so great as
in the effective mass itself.

\section{Conclusions}\label{sec:conclusions}
We have shown how a measurement of the frequency-dependent effective
mass of a granular aggregate, $\tilde{M}(\omega)$, allows us to
predict, accurately, the effects of a grain-filled cavity on the
acoustic properties of a resonant structure.  This fact gives us a
more direct access to an investigation of the underlying physical
mechanisms relevant to the dampening effect of granular media on
structure-borne sound. Crudely speaking, we may think of these
systems as having an effective speed of sound (small) and an
effective viscosity (large).  The dissipation mechanism occurs at
the grain-grain contact level as our simulations have indicated and
our humidity controlled experiments have made unavoidably clear.  As
the humidity is increased there is a large increase in the
attenuation of the fundamental resonance within the grains,
Im($f_g$), which translates to a non-monotonic, but calculable,
variation in the attenuation of the structural resonance in the bar,
Im($f_R$). When the system is taken to a high level of humidity, and
then dried to the same level of humidity as it was at the beginning,
there is a dramatic reduction in attenuation and a dramatic increase
in stiffness of the grain-grain contacts, at the end of this
humid-dry cycle relative to that in the initial dry state.  We
understand this effect in terms of increased solid-on-solid contact
area at the grain-grain contacts.

\begin{acknowledgements} We are grateful to L. McGowan for technical assistance in
collecting the data and to B. Sinha for directing us to the
Timoshenko beam theory.  We are grateful to M. Shattuck for advising
us on the issue of reproducibility of these systems.   We are
grateful for an illuminating conversation with C. W. Frank regarding
the history dependence of the drying effect.  We thank W. K. Martin
for her help with processing the images.  We very much appreciate
several insightful questions from two of the anonymous referees.  We
acknowledge financial support from the US Department of Energy,
Chemical Sciences, Geosciences.
\end{acknowledgements}
\vspace{5mm}

\noindent {\bf Appendix A: Treatment of Bar Resonances Using
Timoshenko Theory}
\renewcommand{\theequation}{A-\arabic{equation}}
\setcounter{equation}{0} \vspace{5mm}

The very simple one dimensional theory of wave propagation in a
flexing bar \cite{Kinsler} is not accurate enough for a quantitative
treatment of the effect of a granular effective mass on the
frequency shift and change of quality factor. Basically this is
because the ratio of length to thickness of our bar is not large
enough  (the frequency is not low enough).  Accordingly we start
with the more sophisticated, but still one-dimensional, theory
developed by Timoshenko \cite{timoshenko}. The equation of motion
for the vertical displacement is \beq{tim}
\begin{array}{l}E I \frac{\partial^4 u}{\partial x^4} + \rho A \frac{\partial^2
u}{\partial t^2} \\
\\
- \rho I \left (\frac{E}{k \mu} + 1 \right ) \frac{\partial^4
u}{\partial x^2
\partial t^2}+ \frac{\rho^2 I}{k \mu} \frac{\partial^4 u}{\partial t^4} =
0\:\:\:\:, \end{array}\eeq where $E$ is the Young's modulus, $A =w
d$ is the cross-sectional area of the bar in terms of its thickness,
$w$, and depth, $d$.  $\rho$ is the density of the bar and $\mu$ is
the shear modulus.  $\rho I$ is the moment of inertia of the
cross-section of the bar relative to its mid-point: $I = (1/12)w^2
A$ for a rectangular bar.  $k$ is a shape parameter which we take to
be equal to (8/9) as is appropriate for a bar of rectangular
cross-section.  The simple theory for a flexing bar corresponds to
keeping only the first two terms on the LHS of Eq. \rf{tim}.

At each frequency, which is in general complex-valued, the solution
can be written as a linear combination of four linearly independent
solutions: \beq{soln} \begin{array}{l}u = [ A \sin(q_1x) + B
\cos(q_1 x) \\
\\
+ C \sinh(q_2 x) + D \cosh(q_2 x)] e^{-i \omega
t}\end{array}\:\:\:,\eeq where $q_{1,2} (\omega)$ are determined by
a direct substitution of Eq. \rf{soln} into Eq. \rf{tim}.  Because
we are interested only in the fundamental mode, which is symmetric
w.r.t. $x=0$, we need only consider the region $0<x<L/2$.  The
coefficients $[A, B, C, D]$ are determined by the requirement that
certain boundary conditions be satisfied at the ends on the bars, $x
= \pm L/2$, and at the center, $x=0$.  However, the coefficients $A$
and $C$ are {\it not} equal to zero, as we shall see, due to the
fact that the slope, $\frac{\partial u}{\partial x}$, is not
continuous at $x=0$.

As we discuss in the main text, we assume the dynamic effective mass
is located at the point $x=0$ and we are treating the bar as
effectively a one-dimensional object.  Taking into account the
missing mass of steel in the cavity the density is \beq{density2}
\rho({ x}) = \rho_0 + \frac{\tilde{M}(\omega)-M_h}{A} \delta({ x })
\eeq where $\rho_0$ is the density of steel, $M_h = \rho_0 \pi a^2
L$ is the missing mass of steel taken from the cavity, and $A$ is
the cross-sectional area of the bar.  One of the boundary conditions
is that the net force exerted by the bar on the cavity must equal
the net mass times the acceleration, viz:
\beq{accbc}\begin{array}{ccl}
-[\tilde{M}(\omega) -M_h] \omega^2 u(0)& =& -[F(0^+)-F(0^-)]\\
&&\\
& =& -2F(0^+)\:\:\: \end{array}\eeq  where $F(x)$ is the force which
the bar exerts on the element $x^+$.  Within the context of
Timoshenko theory, it is specified in terms of specific spatial and
temporal time derivatives of $u(x)$ \cite{timoshenko}.  Thus Eq.
\rf{accbc} gives one of the four necessary boundary conditions on
the coefficients $[A, B, C, D]$.

The presence of the cavity at $x=0$ makes this region of the bar
very bendable.  The radius of curvature in this region is very much
smaller than it is away from the cavity.   We model this effect as
if the left and right halves of the bar are hinged together such
that the bending angle between the two halves is proportional to the
bending moment imposed by the bar: \beq{bm} \frac{\partial
u}{\partial x}\mid_{0^+} = \xi \tau(0^+)\:\:\:,\eeq where $\tau(x)$,
the bending moment, is also given in terms of spatial and temporal
derivatives of $u(x)$. The parameter $\xi$ characterizes the
``bendableness" of the cavity region. $\xi =0 \leftrightarrow
\frac{\partial u}{\partial x}\mid_{0^+} = 0$ corresponds to no
additional extra compliance, whereas $\xi \rightarrow \infty
\leftrightarrow \tau(0^+) =0$ corresponds to a completely flexible
hinge.  We determine the numerical value of $\xi$ by the requirement
that the measured and computed resonance frequency of the bar with
an empty cavity, $\tilde{M} \equiv 0$, must match.  It is the only
free parameter in the theory.

The remaining two boundary conditions are that both the force and
the bending moment vanish at the end of the bar: \beq{fend} F(L/2) =
0\eeq and \beq{Tend} \tau(L/2) = 0 \:\:\:. \eeq.

There are, therefore, four homogeneous linear equations in the four
unknowns ${\bf X} =[A, B, C, D]$ i.e. $H_{ij}X_j = 0$ and so the
determinant of the matrix of coefficients must vanish for any
non-trivial solution.  We search for such  complex-valued roots,
$\omega_R$, such that det$[H(\omega_R)] = 0$ using Muller's method
\cite{numrec}. In general there is a countably infinite number of
such normal mode solutions.  One must choose a set of three initial
frequencies with which the algorithm initiates the search.
Generally, we have chosen the three to be of the form [0.95, 1.00,
1.05]$\omega_S$, where the seed frequency, $\omega_S$, may be
complex-valued.  The normal mode frequency to which the algorithm
converges, if any, depends upon the value of $\omega_S$ that is
chosen.  In the case of the data shown in Figure \ref{glove_box} the
value of the seed frequency needed to be very close to the
resonances around 4 (kHz), but did not need to be very close to the
ones found near 3 (kHz).

As a check, we have compared the theory vs. experiment for a similar
steel bar having no cavity at all: $\tilde{M}\equiv M_h$ and $\xi
\equiv 0$.  The dimensions of this bar are L X W X D = 20.32 X 3.05
X 3.81 cm and we used parameters $\rho = 7830 \;(kg\cdot m^{-3})$, $V_c = 5800\;(
m\cdot s^{-1})$ and $V_s = 3100 \;(m\cdot s^{-1})$ which we have measured ourselves on the
steel samples.  There are two directions in which this bar can flex;
we measured and we computed the fundamental resonance frequency for
each orientation. The results are shown in Table \ref{tab:nohole}.
We see that the predictions of the full theory agree with the
measured values, to within a few tenths of a percent (better than
the accuracy with which the input parameters are known), whereas the
predictions of the simple flex theory differ by as much as 10\%. The
accuracy of the former approach is more than acceptable for our
purposes whereas that of the latter is not.

\begin{table}
\caption{ Comparison of measured and predicted resonance frequencies
for a bar with no cavity (Hz).}

\bigskip

\begin{tabular}{|c|c|c|c|}\hline
&&&\\
 &Measured & Full Theory & Simple Theory \\
 &&&\\
\hline
&&&\\
Flex Horiz. & 3542 & 3529 & 3793 \\
&&&\\
Flex Vert. & 4272 & 4260 & 4741 \\
&&&\\
\hline
\end{tabular}
\label{tab:nohole}
\end{table}
In order to implement the theory when the cavity is filled with a
granular medium we need a means to compute $\tilde{M}(\omega)$ for
complex-valued $\omega$, though the input data is limited to
real-valued $\omega$.  We describe how to do this in Appendix B.

\vspace{5mm}
\noindent {\bf Appendix B: Analytic Continuation of
\mbox{\boldmath{$\tilde{M}(\omega)$}} for Complex Values of
$\bfomega$.}
\renewcommand{\theequation}{B-\arabic{equation}}
\setcounter{equation}{0} \vspace{5mm}

\begin{figure}
\hbox{ \resizebox{6.5cm}{!} { \includegraphics{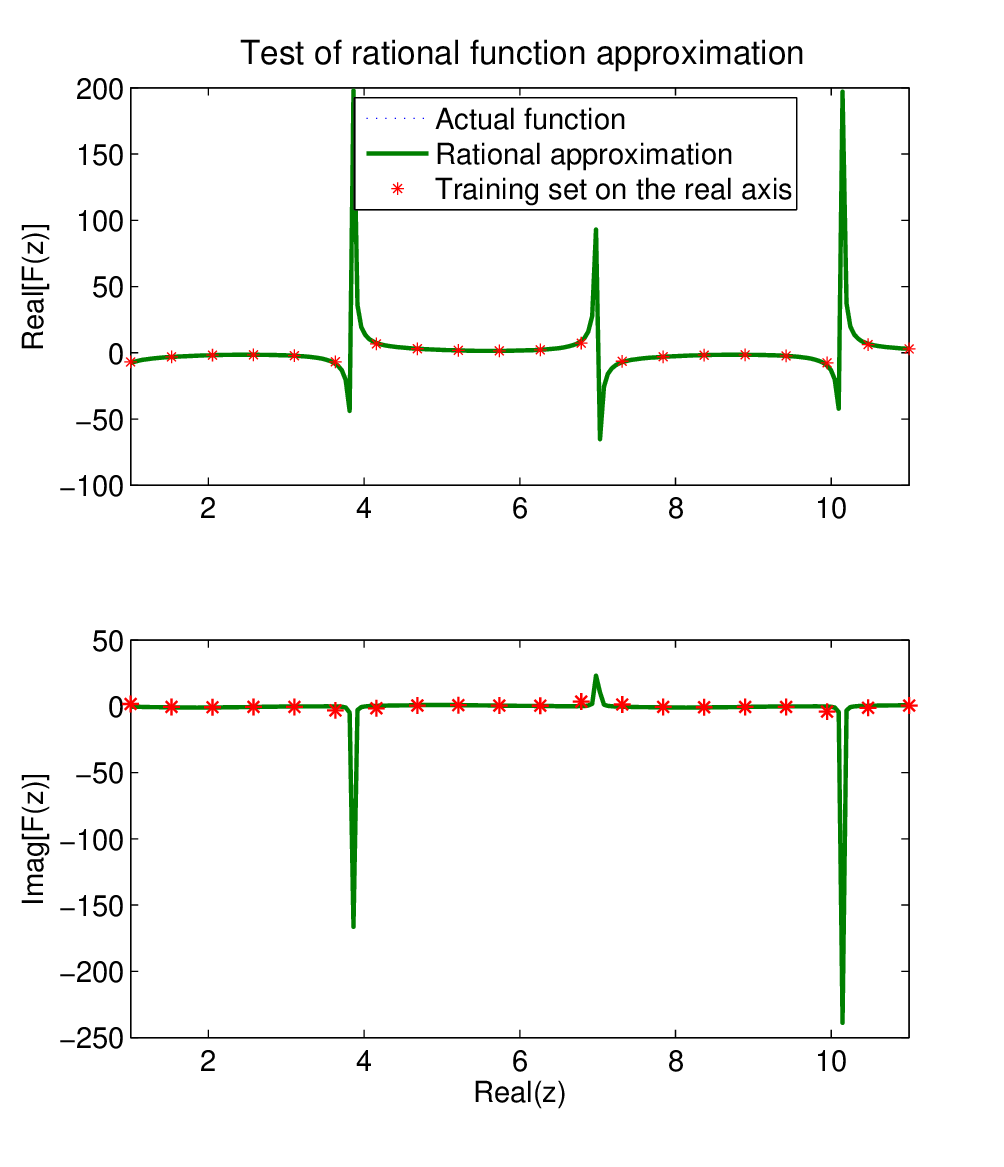}} }
\caption {(Color online) Test of rational function approximation.
The symbols represent the value of the function, Eq. \rf{ztest}, for
values of $z$ on the real axis.  These 20 values are used to
construct a rational function approximation for arbitrary
complex-values $z$. It is compared against the original function for
$z = x-0.095 i$ where we see that there is a near perfect overlay of
the two functions.} \label{test_ratint}
\end{figure}

With the shaker apparatus we have determined the dynamic effective
mass for a series of real-valued frequencies:
$\tilde{M}(\omega_j)\:\:j = 1, \cdots, N$.  In order to use this
information for purposes of computing the resonance frequency in the
flexing bar, as described in Appendix A, we need to be able to
evaluate $\tilde{M}(\omega)$ for arbitrary complex values of
$\omega$. This is because the normal mode frequency of the bar will
be complex-valued, reflecting the attenuation in the system.  We
have chosen to do this analytic continuation using the rational
function technique, a Bulirsch-Stoer algorithm, as described in
Reference \cite{numrec}.  In essence, the algorithm approximates the
function as a ratio of two polynomials - a rational function.  The
algorithm has the property that the rational function passes through
each input datum exactly. Thus, any noise in the input data does
propagate.  In order to test how well this approach actually does
the analytic continuation, we consider the following simple
function: \beq{ztest} F(z) = -\left\{ \frac{1}{\sin[z-(5-i))]} +
\frac{2}{\sin[z-(7-0.1i)]}\right\} \:\:\:.\eeq We have evaluated
this function for 20 different real values of $z$, and these values
are plotted in Figure \ref{test_ratint}.

With these 20 values we construct the rational function
approximation and plot it for values of $z$ on a line in the complex
plane parallel to the x-axis: $z = x-0.095i$. We also plot Eq.
\rf{ztest} on the same graph where we see that the rational function
basically overlays the actual function, even for values of $z$ near
one of the poles.  In fact, the difference between the two never
exceeds 1\%.  This test gives us confidence to use the rational
function to analytically continue $\tilde{M}(\omega)$ into the
complex $\omega$-plane.

\end{document}